\documentclass[aps,pre,twocolumn,superscriptaddress]{revtex4-1} 

\usepackage{graphicx}  
\usepackage{dcolumn}   
\usepackage{bm}        
\usepackage{amssymb}   
\usepackage{amsmath}
\usepackage{siunitx}
\usepackage{physics}
\usepackage{color}
\usepackage{times}
\usepackage{verbatim}

\DeclareMathOperator*{\argmin}{arg\,min} 

\begin{document}

\title{Robust design from systems physics}

\author{Andrei A.\ Klishin}
\affiliation{Department of Physics, University of Michigan, Ann Arbor, MI, 
	USA}
\affiliation{Center for the Study of Complex Systems, University of 
	Michigan, Ann
	Arbor, MI, USA}
\affiliation{John A. Paulson School of Engineering and Applied Sciences, 
	Harvard 
	University, Cambridge, MA 02138, USA}
\author{Alec Kirkley}
\affiliation{Department of Physics, University of Michigan, Ann Arbor, MI, 
	USA}
\author{David J.\ Singer}
\affiliation{Naval Architecture and Marine Engineering, University of 
	Michigan, Ann 
	Arbor, MI, USA}
\author{Greg van Anders}
\affiliation{Department of Physics, University of Michigan, Ann Arbor, MI, 
	USA}
\affiliation{Center for the Study of Complex Systems, University of 
	Michigan, Ann
	Arbor, MI, USA}
\affiliation{Department of Physics, Engineering Physics, and Astronomy, 
	Queen's
	University, Kingston, Ontario, K7L 3N6, Canada}

\email{gva@queensu.ca}

\begin{abstract}
  A crucial challenge in engineering modern, integrated systems is to produce
  robust designs. Ensuring robust design is difficult because
  subsystem couplings produce unpredictable response to
  changes in whole system specifications. Here, we show that the response of
  design elements to whole-system specification changes can be characterized, as
  materials are, using strong/weak and brittle/ductile dichotomies. We find
  these dichotomies emerge from a mesoscale treatment of early stage design
  problems that we cast in terms of stress--strain relationships. Compared
  with other state-of-the-art techniques, we propose a two-factor robustness
  metric that is computable for sets of design solutions. We
illustrate the use of this approach with examples of arrangement problems drawn
from
  naval engineering, however our approach is immediately applicable to a broad
  range of problems in integrated systems design.
\end{abstract}

\flushbottom
\maketitle
%
%
\thispagestyle{empty}

\section{Introduction}
\begin{figure*}
  \begin{center}
  \includegraphics[width=0.9\textwidth]{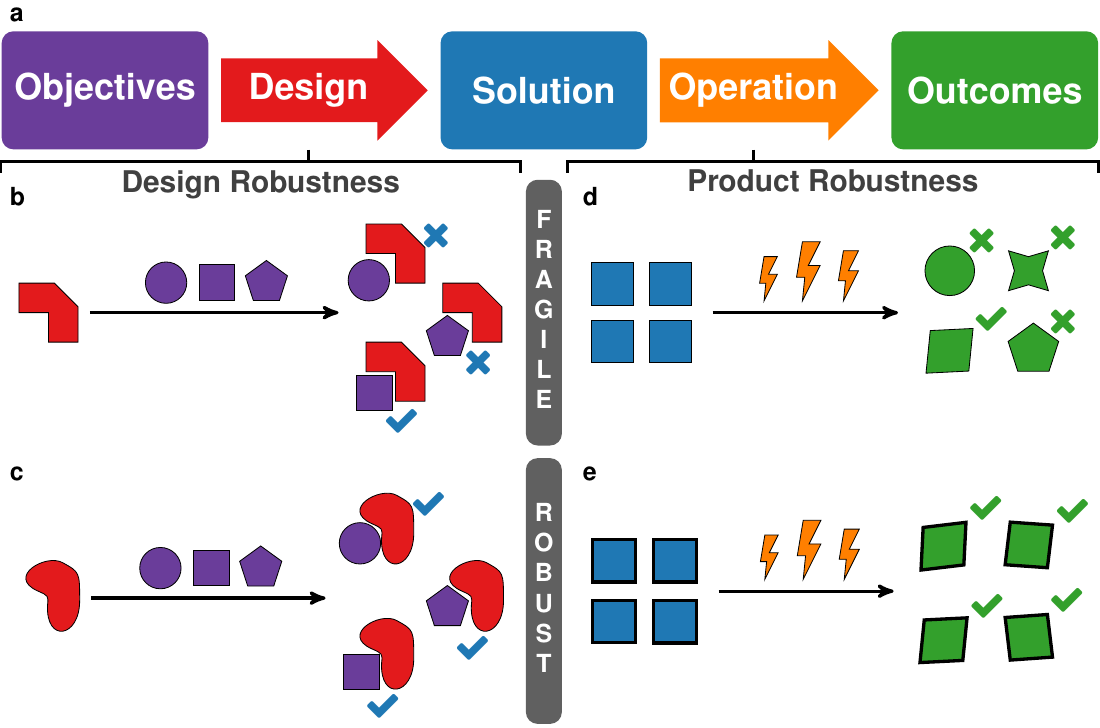}
  \end{center}
  \caption{Schematic representation of the difference between a robust
    design and a robust product.
    (a) The pathway of product design and operation can be represented as a flow
    from a set of objectives, to a design process that produces a solution or
    product, followed by the testing and operation of the product results in one
    or more outcomes. The robustness of the product describes the product's
    performance under different operating conditions. Robust products (panel e,
    blue squares) perform well under different operating conditions (lightning
    bolts), whereas ``fragile'' products (panel d) perform poorly. An analogous
    classification can also be applied to the design process. A robust design
    (panel c) is one in which the same solution (red shape) would be produced to
    meet different sets of objectives (purple shapes), whereas a fragile design
    (panel b) would not stand up to changes in the objectives.
  }
  \label{fig:robust_fragile}
\end{figure*}

\begin{figure*}
  \begin{center}
  \includegraphics[width=0.5\textwidth]{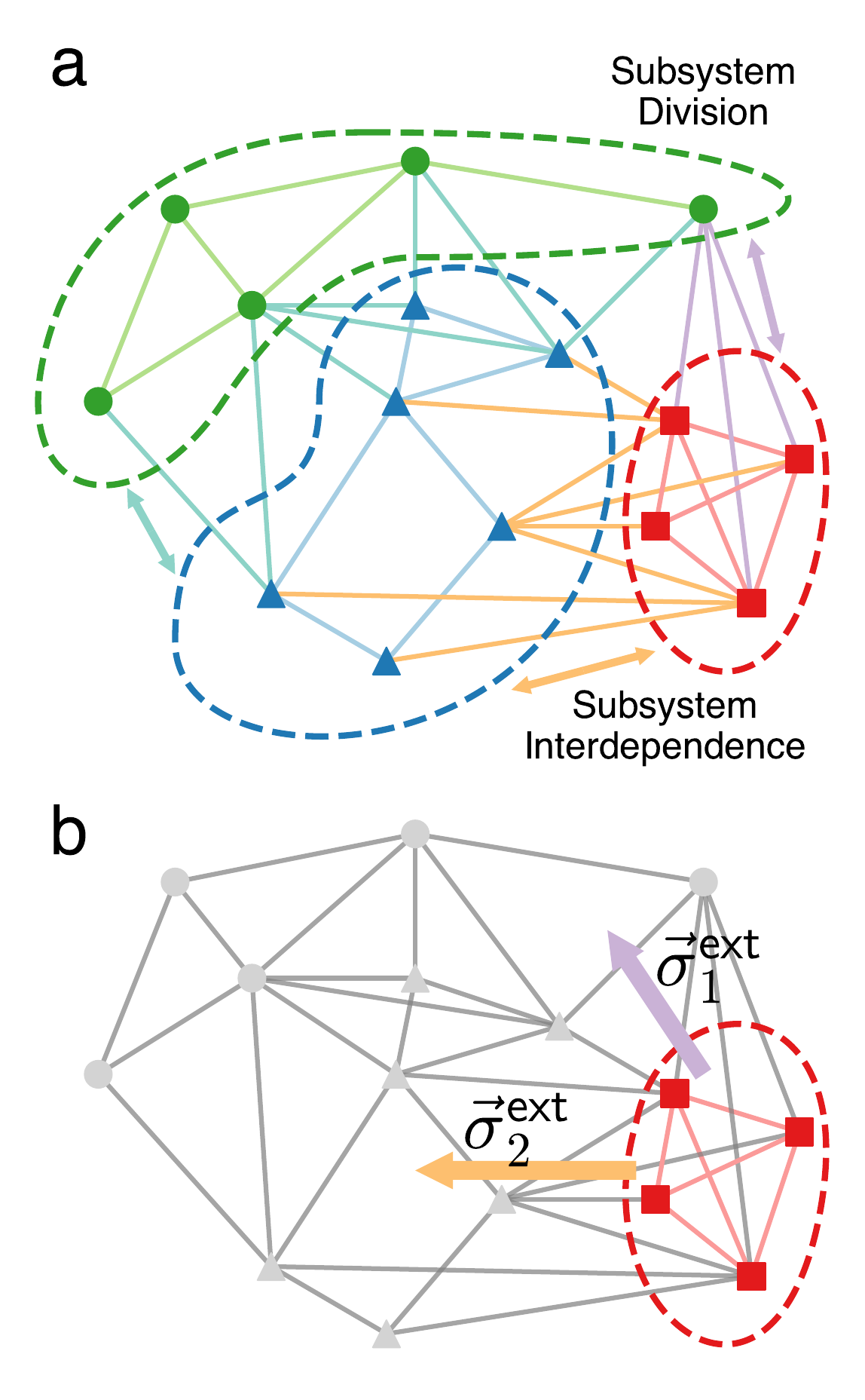}
  \end{center}
  \caption{Schematic representation of interdependencies in a complex,
    integrated system. (a) Complex systems can be comprised of a set of
    connected, interdependent subsystems (distinguished by color). (b) A common
    design problem is to determine the relationship of a component subsystem (in
    red) to the other subsystems (gray). External subsystems induce
    ``design stress'' ($\vec{\sigma}^\textbf{ext}_{1,2}$) on a subsystem that we
    use to characterize robustness.
  }
  \label{fig:schematic}
\end{figure*}

Modern manufacturing and industrial development demand both robust products and
robust designs. Whereas robust products exhibit similar, predictable behavior in
a variety of operating conditions, robust designs preserve design elements under
uncertainty in problem statements (see
Fig.~\ref{fig:robust_fragile}).\cite{taguchi1986introduction,phadke1995quality}
Achieving robust design enhances supply chain stability, avoids rework, and thus
reduces downstream cost and performance
uncertainty.\cite{boehm1988software,love1999determining} Minimizing these
uncertainties through robust design has become both increasingly important, and
increasingly difficult to achieve, as products coming to market incorporate
broader arrays of functionality that rely on the integration of heterogeneous
subsystems.\cite{mckinseymfg} The coupling of heterogeneous subsystems
restricts subsystem component specifications, and small
changes in the design of one subsystem can trigger avalanches of change in
connected subsystems.\cite{Shields2017} Preventing or controlling
avalanches requires developing the ability to not only describe
subsystem interdependencies, but also how interdependencies affect the
robustness of subsystem and overall design.
(see Fig.\ \ref{fig:schematic}).

Throughout engineering, the design of system elements often exploits known
physical phenomena, leveraging knowledge developed through decades or
centuries of
investigation of basic physical science principles. For example, the
principles of robustness of engineering materials have a long
history and a rich language and mathematical apparatus.
In this language, intuitive contrasts such as ``brittle'' vs ``ductile'' or
``strong'' vs ``weak'' achieve precise meaning in terms of performance
thresholds on stress, or localized force, and strain, or localized
displacement.\cite{DieterBacon} Unlike the study of materials, the study of the
basic physical phenomena that underlie the behavior of systems integration is in
its relative
infancy.\cite{Stanley2012failures,buldyrev2010catastrophic,Brummitt2012} Because
of this relative infancy, what it means to be robust, and how to quantify
robustness are open questions.

Here, we operationalize questions about the robustness of subsystem
design via the ``Systems Physics'' approach of Ref.\ \cite{systemphys}. Via
Systems Physics we draw quantitative comparisons between \emph{design} classes,
or architectures, at intermediate, ``mesoscale'' levels of analysis. We find
that at the mesoscale, the robustness of architecture classes can be rigorously
discussed in precisely the same terms that are used to quantify the robustness
of materials, i.e., in terms of stress--strain relationships. The design
stress--strain curves, in the distributed systems we study, generically exhibit
strain softening, i.e.\ a decrease in design stress with increasing design
strain. By, focusing on stress and strain thresholds, we classify the mesoscale
designs by their response to external subsystem coupling as ``brittle'' or
``ductile'' and ``strong'' or ``weak''. We show that the stress--strain
analysis can be concisely summarized in two-factor robustness plots that
directly compare system architectures. For concreteness, we show explicit
examples of brittle, ductile, strong, and weak designs that arise in the context
of a naval architecture-inspired arrangement problem. We show that local
architecture classes can change between brittle and ductile behavior depending
on the form of global design pressure. This analysis of local manifestations of
global design drivers provides a novel form of insight into a ubiquitous set of
challenges faced in industrial design, as well as a new means of communicating
about and achieving robust design.

\section{Robust Design from Statistical Physics}
\label{methods}
Engineering complex systems is a difficult, longstanding problem. Early
systematic design paradigms prescribed optimizing subsystems sequentially
and independently, in the hope of forming a design ``spiral'' that narrows 
to a final, single solution.\cite{evans1959} These ``point-based'' design
approaches rely heavily on mathematical
optimization.\cite{papalambros2000principles} Optimal solutions, however, are
only as good as the underlying models that produce them, and a key source of
model uncertainty is the interaction of the model with external subsystems.
The difficulty of design lies not in the individual subsystems but in their
integration.\cite{chalfant2015} Large systems further compound the potential
for integration failure.\cite{Andrews2012} The failure potential can be
mitigated by focusing not on finding ``good'' solutions, but by focusing on
avoiding ``bad'' ones. Broad-based, so-called ``Set-Based Design'', paradigms
have become influential across automotive,\cite{2004toyota,morgan2006toyota}
aerospace,\cite{bernstein1998} and naval design.\cite{singersbd} Regardless of
the domain, a key challenge in Set-Based Design is to comb through a space of
potential design solutions and eliminate ones that have elements that are
likely to lead to future problems. Those future design problems are likely to
arise when design elements are not robust.

However, describing robustness in set-based and other flexible design
paradigms requires new approaches. Robustness approaches in narrowing,
convergent design \cite{bental2000robust,bertsimas2011theory} describe
single-design solutions, living at a point in design space. In contrast,
set-based design paradigms require determining the robustness not of
\emph{a} design, but of \emph{sets} of designs. The quantitative,
collective treatment of sets is precisely the subject of statistical
physics.

Statistical physics has a long history of describing collective behaviors that
range from the long-known thermodynamics of gases,\cite{gibbs} to more recent
investigations of entropy-driven order,\cite{entint,ordviaent,engent,entchem}
and a host of non-thermal collective phenomena, including flocking
behaviors,\cite{vicsek2012collective} the collective motion of human
crowds,\cite{silverberg2013moshpit} traffic
jams,\cite{chowdhury2000statistical} the synchronization of agricultural
yields,\cite{noble2018spatial} and primate social
dynamics.\cite{flack2017critical} 
A central advantage of statistical physics is its ability to group together
  microscopic system states and investigate the properties of and transitions
  between those groups in the language of free energy (see the conceptual and
  mathematical discussion below). Free energy ideas have been used to classify
  transitions between collective behavior regimes in complex
  systems \cite{koorehdavoudi2016statistical} and cognition.\cite{friston2010free}
Work has also shown\cite{systemphys} that statistical physics approaches can be applied to
systems design problems, under the guise of Systems Physics. Here, we use Systems
Physics to study the intermediate-scale structure of design spaces to study
design robustness.

\subsection{General Approach}
To establish a physics approach for understanding robust systems design,
we take cues from the physics of materials. In materials an
instructive example is a steel rod under mechanical load. Under load the
rod can take one of two qualitatively different states, intact or broken. Before
it breaks, the response of a rod to external forcing can be quantified
using material-dependent relationships between force and deformation, i.e.\
stress--strain relationships. The nature of the stress--strain response of the
rod can be used to concretely describe its material along the independent
axes of \emph{weak--strong} and \emph{brittle--ductile}.\cite{DieterBacon}
Brittle and ductile materials show qualitative differences in behavior.
Both brittle and ductile behavior can, in different industrial contexts, find
appropriate uses. But in either case, determining which material to use requires
knowing how it behaves.

Adapting the materials analogy to systems design requires identifying the
key behaviors and what drives them. In systems design a key factor in robustness
is how design elements behave as they integrate with other subsystems (see
schematic illustration in Fig.~\ref{fig:schematic}a).
Their behavior, in terms of how specifications, locations, etc., respond to
integration is driven by multiple factors. Different subsystems are generally
designed by different designers to satisfy different \emph{design
objectives}. Also generally, each design objective has different relative
importance, which we term \emph{design pressure}. Design pressures act on all
the elements at the same time and thus represent an externally imposed,
whole-system level, or
\emph{global} drive.

Global design pressures manifest themselves \emph{locally} by driving specific 
design elements in different directions, e.g.\ in terms of their physical
locations or specifications. For example, competing design pressures of cost and
performance can produce discord in the specification of design elements.
This discord at the level of elements or subsystems, is analogous to local
mechanical stresses and strains that occur in materials under external load.
This suggests there should be an analogous local measures of force and
deformation, i.e.\ \emph{design stress} and \emph{design strain}, that express
``locally'' how design elements respond to the ``load'' of global design
objectives.

\begin{figure}
  \begin{center}
  \includegraphics[width=0.5\textwidth]{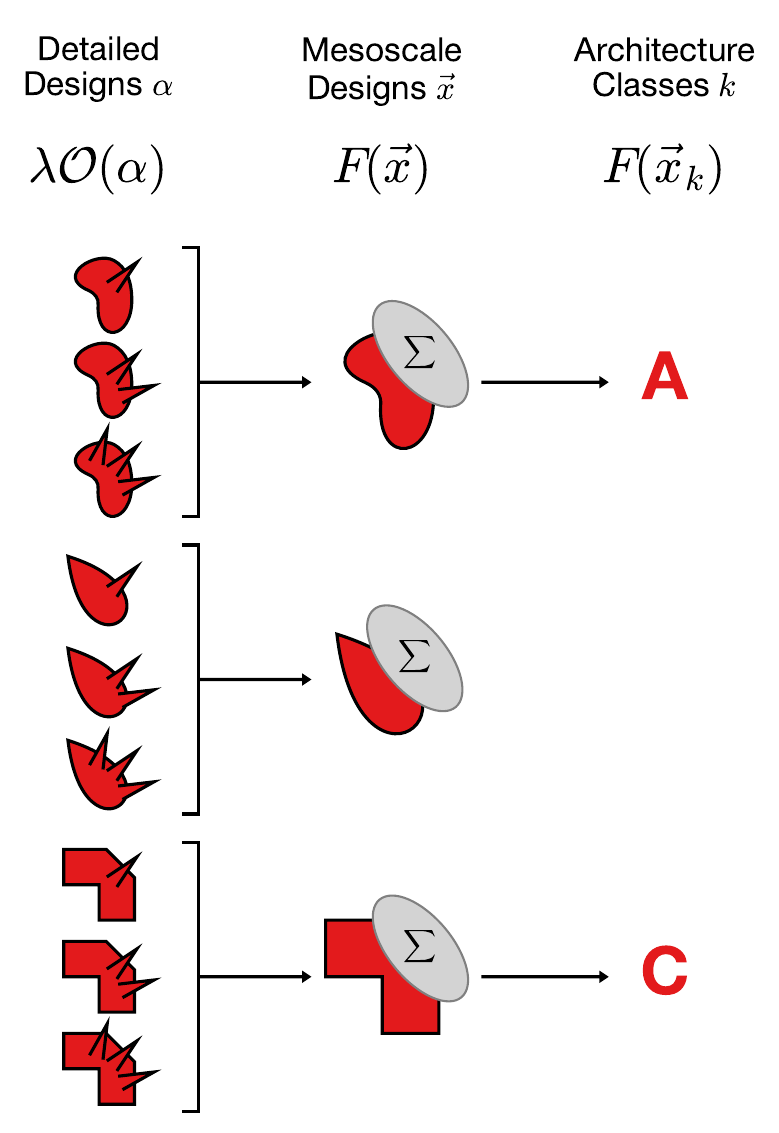}
  \end{center}
  \caption{Detailed designs $\alpha$ group into mesoscale designs $\vec{x}$, and
  locally optimal mesoscale designs form architecture classes $k$. Each detailed
  design is characterized by multiple features (here system shape and spike
  pattern) and quantified by the design objective $\lambda \mathcal{O}(\alpha)$.
  We use the system shape as the feature $\vec{x}$ to define mesoscale designs
  by summing over all spike patterns to get the Landau free energy $F(\vec{x})$.
  Top and bottom shapes are each better than the middle one and thus form
  architecture classes, here $A$ and $C$, in the local free energy minima
  $F(\vec{x}_k)$.}
  \label{fig:mesoscale}
\end{figure}

If global design pressures are connected to local design stress and strain,
how can this be quantified? The challenge in quantifying the global--local
connection is that meeting global design objectives is the collective
result of all component elements. Moreover, when design features could be
placed in a number of possible locations and/or could meet different
specifications, this produces a combinatorial explosion of possible design
states. The challenge of describing the collective behaviors of
combinatorially large numbers of states has an analogue in the thermodynamics of
atomic systems, a problem that prompted the development of statistical physics.
Here, instead of using it to group states of atoms, we will use
statistical physics to group designs.

Our statistical physics approach to quantify design stress and strain
relies on grouping designs that share at least one feature. (see Fig.\
\ref{fig:mesoscale} for an illustration). We term a group of designs with
a shared feature as a \emph{mesoscale} design. Each mesoscale design needs
to be described by a quantity that encodes its properties. The first
quantity that needs to be encoded is the number of designs the mesoscale
design has grouped by common feature. Mesoscale designs that group many
detailed designs need to be distinguished from ones that group few. The
second quantity that needs to be encoded is how well the grouped designs meet
global design objectives. These two factors can be lumped together
into a mathematical function, referred to as a \emph{free energy} in statistical
physics. The free energy of a mesoscale design decreases as
the number of detailed designs that comprise it increases, or as the detailed
designs better satisfy the design objectives, or both. In contrast, changes that
reduce the number of detailed designs or their suitability for the objectives
increase the free energy.

By grouping designs and computing their free energy, mesoscale designs
reduce the complexity of a large number of detailed designs to be reduced to
the consideration of a small number of features of interest. Describing
designs in terms of features of interest has two advantages. First,
features of interest that sit at local minima of the free energy correspond to
locally optimal mesoscale designs. The deviation from a locally
optimal feature set gives a definition of \emph{design strain} and the
free energy change this deviation induces can be used to define \emph{design
stress}. Second, The set of designs that receive a ``pull'' in design
stress toward the same feature set, akin to a watershed, constitute a
``basin'' in design space that can define an\emph{architecture class}.

Grouping designs by feature sets provides each architecture class with a
definition of design stress and strain. Appropriate stress and strain
definitions, in turn, inform a two-factor determination of robustness. To see
why, the materials analogy is again instructive. In materials, robustness is
determined by response to mechanical stress and strain. Mechanical stress and
strain give two key performance indicators of material performance under
different kinds of external influence: maximal loading, or ``ultimate stress'',
and maximum deformation, or ``ultimate strain''. In everyday language, materials
with high ultimate stress are strong and low ultimate stress are weak; materials
with low ultimate strain are brittle and high ultimate strain are ductile. Given
design analogues of material stress and strain, computing the analogous
robustness stress and strain thresholds will provide weak/strong and
brittle/ductile classifications for designs.

\begin{figure*}
  \begin{center}
  \includegraphics[width=0.9\textwidth]{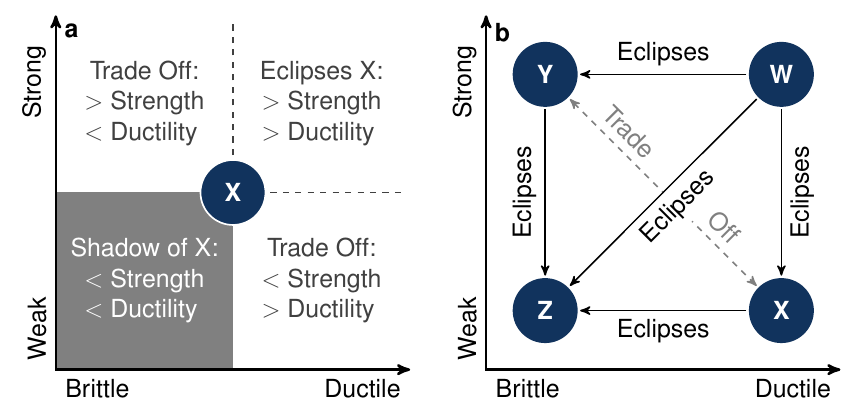}
  \end{center}
  \caption{Two-factor robustness ($R^2$) comparison of design
  architectures.
  Design architectures can be described by their response to external design
  stress (e.g., changes in cost) or design strain (e.g., changes in
  specification limits). The stress--strain response can be used to determine
  the robustness of an architecture, and to compare architectures.
  Panel (a) shows that locating an architecture on robustness axes of
  weak--strong and brittle--ductile facilitates the comparison of that
  architecture $\mathsf{X}$ with other potential architectures. The existence of
  architecture $\mathsf{X}$ casts a ``shadow'' (lower left quadrant) over other
  potential architectures that would be less robust by both measures (lower
  strength, lower ductility).  Conversely situating $\mathsf{X}$ also identifies
  (upper right quadrant) forms of robustness, that if they were found in other
  potential architectures would ``eclipse'' the robustness of $\mathsf{X}$
  (greater strength, greater ductility). The other two quadrants (upper left,
  lower right) describe regions where potential architectures would involve a
  trade-off in forms of robustness (either in strength or ductility) between
  architectures.
  Panel (b) illustrates how this could inform comparison of a set of
  architectures $\mathsf{W}$, $\mathsf{X}$, $\mathsf{Y}$, and $\mathsf{Z}$. In
  this illustration $\mathsf{W}$ eclipses all other architectures either in
  terms of strength ($\mathsf{X}$) ductility ($\mathsf{Y}$) or both
  ($\mathsf{Z}$). If architecture $\mathsf{W}$ did not exist, $\mathsf{Z}$ is
  eclipsed by both $\mathsf{X}$ (in ductility) and $\mathsf{Y}$ (in strength)
  but a choice between $\mathsf{X}$ and $\mathsf{Y}$ means a trade-off between
  the two forms of robustness.
  }
  \label{TwoAxes}
\end{figure*}

Knowing how weak/strong or brittle/ductile particular design architecture
is is useful. However, it is also important to compare the robustness of
designs. To make this comparison, the existence of weak/strong and brittle/
ductile contrasts suggests plotting architectures on two axes that run
weak--strong and brittle--ductile. A sketch of this is given in
Fig.\ \ref{TwoAxes}a. An architecture $\mathsf{X}$ can be located on a pair of
axes representing the two measures of robustness. The region around $\mathsf{X}$
can be divided into quadrants. Additional architectures would fall into one of
those quadrants, permitting a direct comparison of robustness. We refer to the
lower left quadrant as the shadow of $\mathsf{X}$ because architectures falling
in that region would be inferior to $\mathsf{X}$ in both strength and ductility.
In contrast $\mathsf{X}$ would be in the shadow of any architecture that falls
in the upper right quadrant, because that architecture would have greater
strength and ductility. We refer to that region as the eclipsing region of
$\mathsf{X}$. The other two quadrants, in the upper left and lower right, are
regions where architectures would involve trade-offs with $\mathsf{X}$, either
greater strength but reduced ductility, or greater ductility but reduced
strength.

An example robustness comparison between is sketched in Fig.\
\ref{TwoAxes}b. Fig.\ \ref{TwoAxes}b gives an example of a two-factor robustness
plot ($R^2$-plot) for four architectures, which we label $\mathsf{W}$,
$\mathsf{X}$, $\mathsf{Y}$, and $\mathsf{Z}$. In the scenario depicted in the
$R^2$-plot \ref{TwoAxes}b, $\mathsf{W}$ has the same ductility as
$\mathsf{X}$, but $\mathsf{W}$ is stronger, so all else being equal the
architecture $\mathsf{W}$ would represent a more robust choice. Similarly,
architectures $\mathsf{W}$ and $\mathsf{Y}$ have the same strength, but
$\mathsf{W}$ is more ductile, so all else being equal $\mathsf{W}$ would
represent a more robust choice. Similar reasoning also indicates that
$\mathsf{W}$ is more robust than $\mathsf{Z}$ in both strength and ductility. If
architecture $\mathsf{W}$ was unavailable, similar considerations would
make $\mathsf{Z}$ a less robust choice than either $\mathsf{X}$ or $\mathsf{Y}$.
Comparing $\mathsf{X}$ and $\mathsf{Y}$ shows that the two architectures have
different forms of robustness, $\mathsf{X}$ is more ductile, but $\mathsf{Y}$ is
stronger. In this case, all else being equal, the designer would need additional
information to determine whether strength or ductility is likely to be the more
important measure of robustness.

Fig.\ \ref{TwoAxes} illustrates what could occur in a design process under fixed
external design pressure. However, changes in design pressure can change the
robustness of architectures, echoing e.g.\ the brittle--ductile transitions that
occur in industrial materials \cite{horii1986brittle} and
geology.\cite{byerlee1968brittle} Similar to the varied industrial uses of both
brittle and ductile materials, we anticipate the usefulness of architecture
classes manifesting different forms of robustness. In evaluating robustness, our
goal is to generate knowledge about the possible emergent behaviors in the
design space to inform the human designer, who would make the final design
choice.

\subsection{Systems Physics}
In this section we cast the foregoing general approach into a concrete
mathematical form. Our mathematical model expands upon Systems Physics, a
statistical mechanics framework for design problems introduced in Ref.\
\cite{systemphys}. We consider a design problem with a combinatorially large
ensemble of candidate \emph{detailed design} solutions $\{\alpha\}$. For each
detailed design $\alpha$ we compute several quantitative \emph{design
objectives} $\mathcal{O}_i$ , where $i$ is an index. An example of a
numerical design objective would be the cost of routing a cable between two
functional units; we explain the design objectives for our model system in the
next section. Given a set of design objectives, a standard calculational device
from statistical mechanics that is applied in analogous problems is to associate
an expected average outcome $\left<\mathcal{O}_i\right>$ with each objective.
Given only the above data, information theory implies that the minimally biased
(maximal entropy \cite{jaynes1}) probability distribution for choosing a
detailed design $\alpha$ that matches the objectives with their outcomes is
given by
\begin{equation} \label{eqn:palpha}
p(\alpha)=\frac{1}{\mathcal{Z}}e^{-\sum\limits_{i}\lambda_i
\mathcal{O}_i(\alpha)} \; ,
\end{equation}
where $\mathcal{Z}$ is a normalization constant. High probability designs are
the ones that best fulfill the competing design objectives. Whereas the design
objectives $\mathcal{O}_i$ assumed to be known \emph{a priori} and represent
\emph{what} criteria need to be considered by the designer, the design pressures
$\lambda_i$ represent \emph{how much} each criterion matters, and the choice of
these pressures could differ between stakeholders with different concerns, e.g.\
cost versus performance. Selecting particular values of $\lambda_i$
significantly shapes the probability distribution $p(\alpha)$ and can radically
and abruptly change the types of the preferred designs $\alpha$. Predicting and
quantifying the preferred designs within the ensemble is a goal of Systems
Physics.\cite{systemphys}

The properties of the whole ensemble are contained in the normalization of 
Eq.~\eqref{eqn:palpha} that can be computed as
\begin{equation}\label{eqn:genZ}
  \mathcal{Z}=\sum_{\alpha}e^{-\sum\limits_{i}\lambda_i \mathcal{O}_i(\alpha)}
  \; ,
\end{equation}
and has the familiar form of a partition function from statistical physics. In
statistical physics, the partition function encodes the statistical averages of
design objectives across the whole ensemble.\cite{systemphys} However, summing
over the whole ensemble masks the fact the many detailed designs $\alpha$
achieve the same design objective value $\mathcal{O}_i$, and thus obscures the 
intermediate scale design drivers. Discovering these design drivers requires
studying designs at a higher level of granularity. This granularity is
  given by the architecture classes we discussed in general terms above, and
which we will now define more precisely.

\subsection{Architecture Classes}
We achieve a higher level of granularity by selecting a design feature
$\vec{x}_\alpha$ that multiple detailed designs $\alpha$ share. An example of a
shared feature could be the spatial location of a particular functional
unit or one of its internal operational parameters (e.g.\ pressure
or voltage).  Regardless of
the specific feature chosen, a set of designs sharing a common feature $\vec{x}$
can be described by a \emph{mesoscale design} (see Fig.~\ref{fig:mesoscale}).
The feature space $\{\vec{x}\}$ is typically much smaller than the set of
detailed designs $\{\alpha\}$, but can be used to recover the statistical
information of the full design ensemble via the expression
\begin{equation}
\mathcal{Z}=\sum_{\vec{x}} e^{-F(\vec{x})} \; ,
\end{equation}
where $F(\vec{x})$ is the \emph{free energy}. The free energy quantifies the
effective design objective of the mesoscale design $\vec{x}$, and is determined
by the expression
\begin{equation}
e^{-F(\vec{x})}=\sum\limits_{\alpha}\delta(\vec{x}-\vec{x}_\alpha) 
e^{-\sum\limits_{i}\lambda_i
\mathcal{O}_i(\alpha)} \; .
\label{eqn:freeenergy}
\end{equation}
Here $\delta(\vec{x}-\vec{x}_\alpha)$ is an indicator function, equal to 1 when
the detailed design $\alpha$ belongs to the mesoscale design $\vec{x}$ and 0
otherwise. $F(\vec{x})$ is an example of a so-called Landau free energy
\cite{goldenfeld}, and provides a mesoscale characterization of classes of
designs that share characteristics specified by $\vec{x}$. The procedure of
going from a large ensemble of detailed designs $\alpha$ to mesoscale designs
$\vec{x}$ is known as \emph{coarse graining} in statistical
physics.\cite{goldenfeld} Coarse graining applied to a model system and the
resulting free energy landscape are illustrated in Fig.~\ref{fig:system_setup}.

The shape of the free energy landscape $F$ indicates the relative preference
between different mesoscale designs. Designs that accord better with the balance
of design pressures $\lambda_i$ have smaller $F$, and vice versa. Local minima
of $F(\vec{x})$ are ``best-in-class'' designs. We denote these designs as
$\vec{x}_k$, and index them as $k\in\{A,B,C,\dots\}$ (colored circles in
Fig.~\ref{fig:system_setup}b). However, although best-in-class designs are
defined to be those that best meet a fixed set of design objectives, changes in
the specification of the objectives can alter the classification, so
understanding the robustness of designs is crucial.

\subsection{Quantifying Robustness}
A deviation from a local minimum $\vec{x}_k$ within the feature space gives a
design strain $\vec{\epsilon}_k=\vec{x}-\vec{x}_k$. In design strain
coordinates, design stress is given by the free energy gradient
$\vec{\sigma}(\vec{\epsilon}_k)=-\vec{\nabla}F(\vec{x}_k+\vec{\epsilon}_k)$.
Sufficiently close to the local minimum, design stress pulls the design back to
the minimum, i.e.\ $\vec{\sigma}\cdot\vec{\epsilon}_k<0$. However, at larger
strains in a particular direction, the design can reach a threshold, or saddle
point, in free energy and get pulled by the local design stress to a different
minimum. We call that point the \emph{ultimate strain} and formally define it as
\begin{equation}
  \vec{\epsilon}_k^{\;\textsf{ult}}=\argmin_{\vec{\epsilon}_k}
  \left|\vec{\epsilon}_k\right|:\; \vec{\sigma}\cdot\hat{\epsilon}_k>0 \; ,
  \label{eqn:ult_strain}
\end{equation}
where $|\cdot|$ denotes a suitable vector norm (here we use standard Euclidean
norm) and $\hat{\epsilon}_k$ is a unit vector pointing along the strain
direction. The operator $\argmin$ finds the closest saddle point but still
returns the vector $\vec{\epsilon}_k^{\;\textsf{ult}}$ rather than just its
norm. We illustrate the path from local free energy minima to the saddle points
in an example system in Fig.~\ref{fig:strstr_panels}a,c,e.

As a mesoscale design is strained from $\vec{0}$ to
$\vec{\epsilon}_k^{\;\textsf{ult}}$, it will develop design stress.  To analyze
the stress response, it is convenient to compute the projection of the stress
along the strain direction, $\sigma=\left|\vec{\sigma}(\vec{\epsilon}_k)\cdot
\hat{\epsilon}_k\right|$. From this projection it is possible to compute the
\emph{ultimate stress}, i.e.\ the magnitude of externally exerted stress that
causes designs to switch between classes. Formally, this is given by
\begin{equation}
\sigma_k^\textsf{ult}=\max\limits_{a\in [0,1]}
\left|\vec{\sigma}(\vec{x}_k +a 
\vec{\epsilon}_k^{\;\textsf{ult}})\cdot \hat{\epsilon}_k^\textsf{ult}\right| \; ,
\label{eqn:ult_stress}
\end{equation}
where $a$ is an auxiliary variable parametrizing a straight line from $\vec{0}$
to $\vec{\epsilon}_k^{\;\textsf{ult}}$. While the free energy minimum
$\vec{x}_k$ defines the ``best-in-class'' mesoscale design, the basin of all
mesoscale designs that design stress brings back to the best-in-class design
defines an \emph{architecture class}.

We have defined architecture classes thus far for isolated subsystems. When
subsystem designers incorporate effects that arise from coupling to other
subsystems, other subsystems exert \emph{external} design stress or strain on
the subsystem of interest. External stress and strain correspond, respectively,
to what are referred to in statistical mechanics as ``intensive'' (size
independent) or ``extensive'' (size dependent) modifications of the
specification of the system.

Going back to the analogy of a metal rod, this nomenclature reflects that
a mechanical load can be applied to the rod with two protocols.  One is to
subject the rod to a fixed external stress force, or intensive modification, and
measure the resulting strain.  The other is to subject the rod to a fixed linear
strain, or extensive modification, in form of stretching or compression and
measure the resulting stress.

Protocols for materials response have direct analogues in systems design.
In systems design, an example of external design stress would arise from the
need to route a connection from a functional unit to an external subsystem, with
the direction and cost per unit length specified for the connection. This
scenario creates a uniform design stress $\vec{\sigma}^\textsf{ext}$ on the
subsystem (Fig.~\ref{fig:system_setup}a), and the new local optimum would be
found at the location where the internal design stress balances the external
$\vec{\sigma}^\textsf{ext}+\vec{\sigma}=0$. An example of external strain would
be the need to position an additional object in the location $\vec{x}_k$, thus
requiring the shift of subsystem design features by design strain of
$\vec{\epsilon}{\,}^\textsf{ext}$ away from the minimum. In either case,
external stress or strain may or may not push the mesoscale design into a
different architecture class basin. Resisting the architecture class shift is
the property that we call \emph{robustness}.

We determine the robustness of each architecture class by computing
the design stress--strain curves. From these curves, we extract the
ultimate stress and strain for each architecture class and plot them together
without averaging in Fig.~\ref{fig:ultimate_strstr}. These stress--strain
relationships facilitate the characterization of each design architecture class
as weak or strong by comparing the respective $\sigma_k^\textsf{ult}$ among
different architecture classes $k$. Weak designs have small
$\sigma_k^\textsf{ult}$, whereas strong
designs have large $\sigma_k^\textsf{ult}$. We also characterize designs as
brittle or ductile by comparing the relative
$|\vec{\epsilon}_k^{\;\textsf{ult}}|$. Brittle designs have small
$|\vec{\epsilon}_k^{\;\textsf{ult}}|$, whereas ductile designs have large
$|\vec{\epsilon}_k^{\;\textsf{ult}}|$. Considering both strength and ductility
gives us the two-factor robustness, $R^2$, of the architecture classes,
presented in Fig.~\ref{fig:twofactor}.

\subsection{Beyond Ultimate Stress}
Whereas the determination of ultimate strain can characterize the robustness of
an architecture class, it is also important to understand what happens once
architecture classes are pushed beyond their viability limit. Doing so requires
understanding the configuration of architecture classes under large external
stress.

To model the external stress, we consider not only the free energy of the
subsystem of interest $F(\vec{x})$ that depends on the design feature
$\vec{x}$, but also the free energy of an external subsystem
$F'(\vec{x}\,')$ that depends only on the design features $\vec{x}\,'$ of the
external subsystem. The interdependence of the two subsystems is captured by the
interaction free energy $F_{int}(\vec{x},\vec{x}\,')$. The goal is to 
understanding what happens in the system of interest, under the assumption
the external subsystem can take any configuration it prefers. We
carry this out by integrating out the external subsystem, leaving a
description of the remaining subsystem of interest. Mathematically, the
procedure is similar to
to the earlier coarse-graining procedure in Eq.\ \eqref{eqn:freeenergy},
  \begin{equation}
e^{-\tilde{F}(\vec{x})}=\sum\limits_{\vec{x}\,'} e^{-F(\vec{x})-F'(\vec{x}\,')-F_{int}(\vec{x},\vec{x}\,')}.
\label{eqn:Ftilde}
\end{equation}

In general, performing the computation in Eq.\ \ref{eqn:Ftilde} is
challenging, since it requires a detailed model of the external subsystem.
However, we note that free energy is only defined up to an additive constant
which does not affect the locations or properties of local minima. Thus we
can get insight into the effect of external couplings by adopting a simplified
form of the interaction free energy:%
\begin{equation}
  F_{int}(\vec{x},\vec{x}\,')\approx
  (\vec{x}\,'-\vec{x})\cdot \vec{\sigma}^\textsf{ext}+\text{const.}
\end{equation}
This form of interaction free energy describes a uniform external design
stress $\vec{\sigma}^\textsf{ext}$ applied to each feasible design in the
domain of design feature $\vec{x}$, caused for example by the addition of a
cable of fixed direction and cost per unit length. Computationally, this form
of interaction makes the summation in Eq.\ \ref{eqn:Ftilde} separable and
gives the effective free energy landscape as:%
\begin{equation}
  \tilde{F}(\vec{x})=F(\vec{x})-\vec{x}\cdot\vec{\sigma}^\textsf{ext}+\text{const.}
\end{equation}

In general, $\tilde{F}$ and $F$ will have different sets of local minima, i.e.\
design configurations that are best-in-class will change in the presence of
external stress. We investigate the effect of variable external stress by
considering different vectors $\vec{\sigma}^\textsf{ext}$ that span the space
$\{\vec{\sigma}^\textsf{ext}\}$. If the external stress
$\vec{\sigma}^\textsf{ext}$ is much
larger than any internal design stresses $-\vec{\nabla}F(\vec{x})$ naturally
arising in a given architecture class, the subsystem is completely dominated by
external stress that eliminates candidate architecture classes, reducing design
richness.

We characterize the loss of design richness under external stress by finding the
domain in $\{\vec{\sigma}^\textsf{ext}\}$ space in which a minimum of the same
type $k$ exists. Here, by ``same'' we mean a minimum that moved less than some
threshold $\Delta x^\textsf{th}$ under a small change of stress
$\delta\vec{\sigma}^\textsf{ext}$
\begin{equation}
\left|\vec{x}_k(\vec{\sigma}^\textsf{ext}+\delta\vec{ 
\sigma}^\textsf{ext})-\vec{x}_k(\vec{\sigma}^\textsf{ext})\right|<\Delta x^\textsf{th}\;.
\label{eqn:minima_class}
\end{equation}
We illustrate how external stress affects the viability ranges of subsystem
design classes in Fig.~\ref{fig:minima_diagram} using Venn diagrams in the
$\{\vec{\sigma}^\textsf{ext}\}$ space. Together, the viability ranges for
architecture classes and the analysis of their ultimate design stress and strain
constitute the quantitative knowledge required for robust system design.

\section{Example System: Model, Results, and Discussion}
\subsection{Model System}
\begin{figure*}
  \begin{center}
  \includegraphics[width=0.9\textwidth]{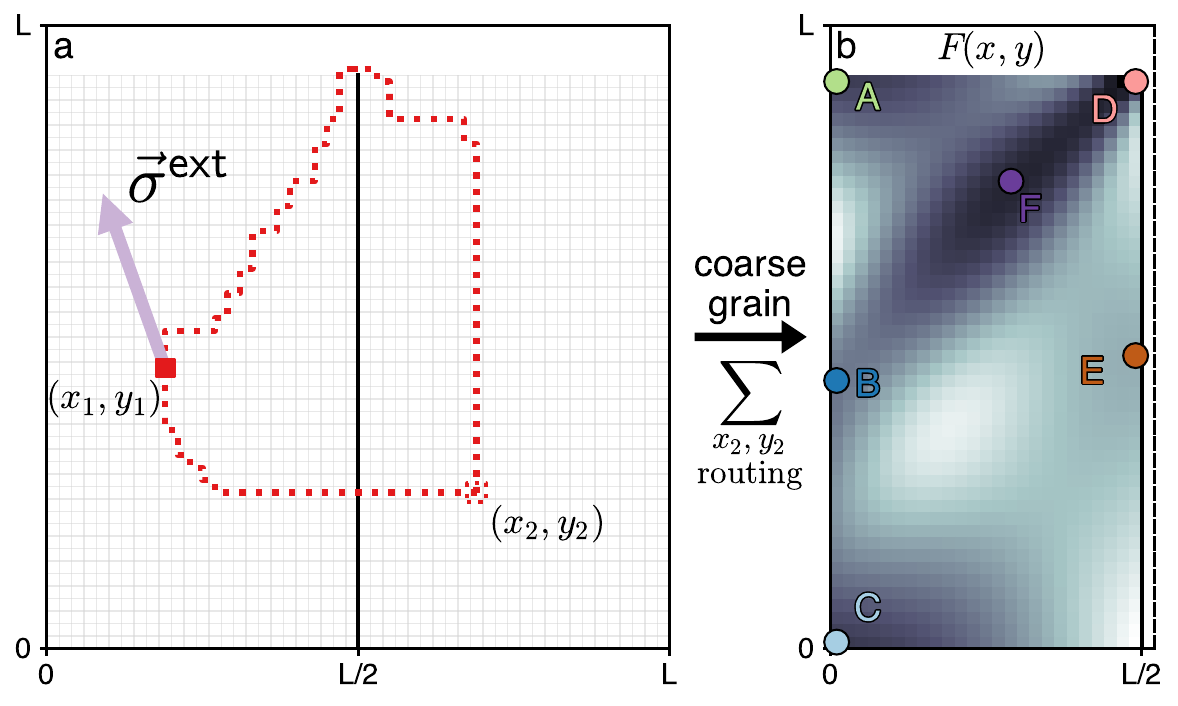}
  \end{center}
  \caption{Schematic representation of the specific subsystem investigated for
  robustness. (a) Two connected functional units are placed in positions
  $(x_1,y_1)$ and $(x_2,y_2)$. There are two qualitative ways to connect them
  along the shortest Manhattan route: either directly by drilling a hole through
  the bulkhead, or by first routing up to the bulkhead, over it and back down.
  The position of the first unit is taken as the design feature $\vec{x}$, while
  the position of the second one and the possible routings are ``integrated
  out'' by computing the free energy via Eqn.~\ref{eqn:free_energy_xy}. The
  external design stress on the system has the form of a constant force
  $\vec{\sigma}^\textsf{ext}$ shown with a purple arrow. (b) Coarse graining
  procedure leads to the free energy landscape $F(x,y)$ for the possible
  positions of the first unit in the part of the domain left of the bulkhead.
  Local free energy minima are identified with architecture classes labelled with
  capital letters $A$ through $F$ and distinct colors.}
  \label{fig:system_setup}
\end{figure*}

The Systems Physics based robustness analysis developed in the previous section
can be applied to a broad range of design problems. For concreteness, we
will illustrate its use in
arrangement problems that arise in Naval Architecture, taking as a specific
example an established model of early stage ship design.\cite{Shields2017} Our
ship design model involves embedding a network of many functional units into a
ship hull of fixed geometry, and routing connections between the units.
Large-scale effects on unit routing induced by changes in design pressure were
studied in Ref.~\cite{systemphys}. Here, we study this model at the mesoscale to
understand the robustness of system design. To determine robustness, we focus on
a subsystem with two connected functional units that is externally connected to
other subsystems, modelled via external design stress (see
Fig.~\ref{fig:system_setup}a).

The subsystem is situated in a square domain of $L\times L$ discrete cells. The
domain is separated along the middle line into two compartments housing one
functional unit each, divided up to height $h_\textsf{bh}$ by a watertight
bulkhead. The bulkhead serves to prevent simultaneous water flooding of both
compartments in case one of them is breached. It is possible to drill a hole
through the bulkhead, reinforce that hole, and route a connection through it.
However, such a hole bears risks that affect ship survivability. To assess
how survivability considerations affect the routing problem, we consider two
distinct types of routing: either along the shortest possible route through
the bulkhead, or along the shortest possible route around the top of the
bulkhead. The connections are only routed horizontally and vertically, so there
is a large but finite number of possible routings for each choice of positions
of the two units. A particular realization of unit positions and routings of two
types is shown in Fig.~\ref{fig:system_setup}a. The possible positions of the
two units $(x_1,y_1)$ and $(x_2,y_2)$ along with the choices of particular
routing form the detailed design space $\{\alpha\} =
\{(x_1,y_1,x_2,y_2,\text{routing})\} $.

Within the design space, detailed designs are evaluated with respect to two
design objectives, $\mathcal{O}_1,\mathcal{O}_2$, and corresponding design
pressures $\lambda_1,\lambda_2$
\begin{align}
\mathcal{O}_1\equiv& E=C\left( |\Delta x|+|\Delta y| \right), & \lambda_1&\equiv 1/T ;
\nonumber\\
\mathcal{O}_2\equiv& B, & \lambda_2&\equiv \gamma \; .
\end{align}
The first design objective $E$ represents the cost of the routing, linearly
proportional to the Manhattan (taxicab/grid) length of the routing used. The
corresponding design pressure is \emph{inverse cost tolerance} $T$, similar to
the thermodynamic temperature. Low cost tolerance means that designs with
shorter routings are strongly preferred, whereas high cost tolerance means that
cost is not a strong factor in choosing a design. The second design objective
$B\in \{0,1\}$ is a binary variable indicating whether a given design routes
through the bulkhead (1) or goes around (0). The corresponding design pressure
is the bulkhead penalty $\gamma$ that quantifies the survivability penalty
associated with routing through the bulkhead. Low, near-zero, values of $\gamma$
mean that routing through or around the bulkhead are equally preferable, whereas
high values of $\gamma$ strongly suppress routing through the bulkhead.

In terms of these specific design objectives and design pressures, the partition
function \eqref{eqn:genZ} takes the form
\begin{equation}
  \mathcal{Z}=\sum_{\alpha}e^{-\frac{E(\alpha)}{T}-\gamma B(\alpha)} \; ,
  \label{eqn:modelZ}
\end{equation}
where the sum over $\alpha$ runs over \emph{the whole set} of possible detailed
designs. To group the detailed designs into mesoscale designs, we use
the position of the \emph{left} functional unit $\vec{x}=(x_1,y_1)$ as the
design feature of interest. The position of the right functional unit
$(x_2,y_2)$ and the routings are integrated out. The position of the left unit
then has the associated free energy $F(x,y)$
\begin{equation}
e^{-F(x,y)}=\sum\limits_{\alpha}\delta(\vec{x}-\vec{x}_1(\alpha))e^{-\frac{E(\alpha)}{T}-
\gamma 
B(\alpha)} \; .
\label{eqn:free_energy_xy}
\end{equation}

An example of the free energy landscape is shown in
Fig.~\ref{fig:system_setup}b. Local minima of the free energy are associated
with the architecture classes $A$ through $F$. We find that the free energy
landscape and the pattern of architecture classes vary greatly with the design
pressures $T,\gamma$, and that in all cases the design robustness is given
directly by Eqns.~\ref{eqn:ult_strain}-\ref{eqn:ult_stress}.  There are several
large-scale reorganizations between architecture classes as the design pressures
$T$ and $\gamma$ are varied. These reorganizations are analogous to phase
transitions in thermodynamic systems, but are not sharp transitions because the
design problems we study have finite size. Despite the finite sizes, it is
possible to approximately determine where the reorganization of architectures
occurs.\cite{systemphys} In this model reorganization occurs around a \emph{cost
phase transition}: at low $T<T_\textsf{crit}\sim C/\ln 2$ units prefer short
connections to minimize the routing cost, whereas at large $T>T_\textsf{crit}$
they prefer long connections to maximize the flexibility in carrying out the
routing.\cite{systemphys} Another large-scale reorganization is the transition
of the average bulkhead penetration $\left< B \right>$, or fraction of designs
routing through the bulkhead: it approaches 0 for simultaneously large $T$ and
$\gamma$ (preferring flexibility in routing and suppressing bulkhead
penetration), and approaches 1 when either $T$ or $\gamma$ is small (preferring
low-cost routing and allowing bulkhead penetration, or a combination of both).
As we will find below, the origin of these large-scale reorganizations can be
traced in the mesoscale through the appearance and disappearance of architecture
classes and changes in their robustness.

To capture the reorganization of architecture classes, we study the routing
problem for a range of design pressures $\{T,\gamma\}$. We fix our system of
units by setting $C=1.0$, which fixes the units of cost tolerance $T$. To
maximize illustrative power, we seek a set of choices of $\gamma$ and $T$
that allows for the study of all the possible architectural organizations
with the fewest number of state points.  For this purpose, a suitable
choice is to scan along the line of $\gamma=2.0$ with $T\in [0.5,2]$
because that choice crosses both the cost- and bulkhead penetration-based
reorganizations. We also make specific choices of system geometry parameters.
Since the positions of the functional units are discrete,
we cannot reliably resolve the stress--strain curves on length scales less than
1 cell. Moreover, the comparison of robustness between architecture
classes requires the system domain to be sufficiently large to
support multiple architecture classes. To this end, we set the domain size
$L=50$, with the bulkhead going up to $h_\textsf{bh}=46$, though our analysis
and results are similar for different system sizes. We find that ultimate
strains vary from 1.5 to 18 cells, allowing us to reliably distinguish the
architecture classes along both weak--strong and brittle--robust axes.

\subsection{Results and Discussion}
\subsubsection{Design Stress and Strain}
\begin{figure*}
\begin{center}
\includegraphics[width=0.7\textwidth]{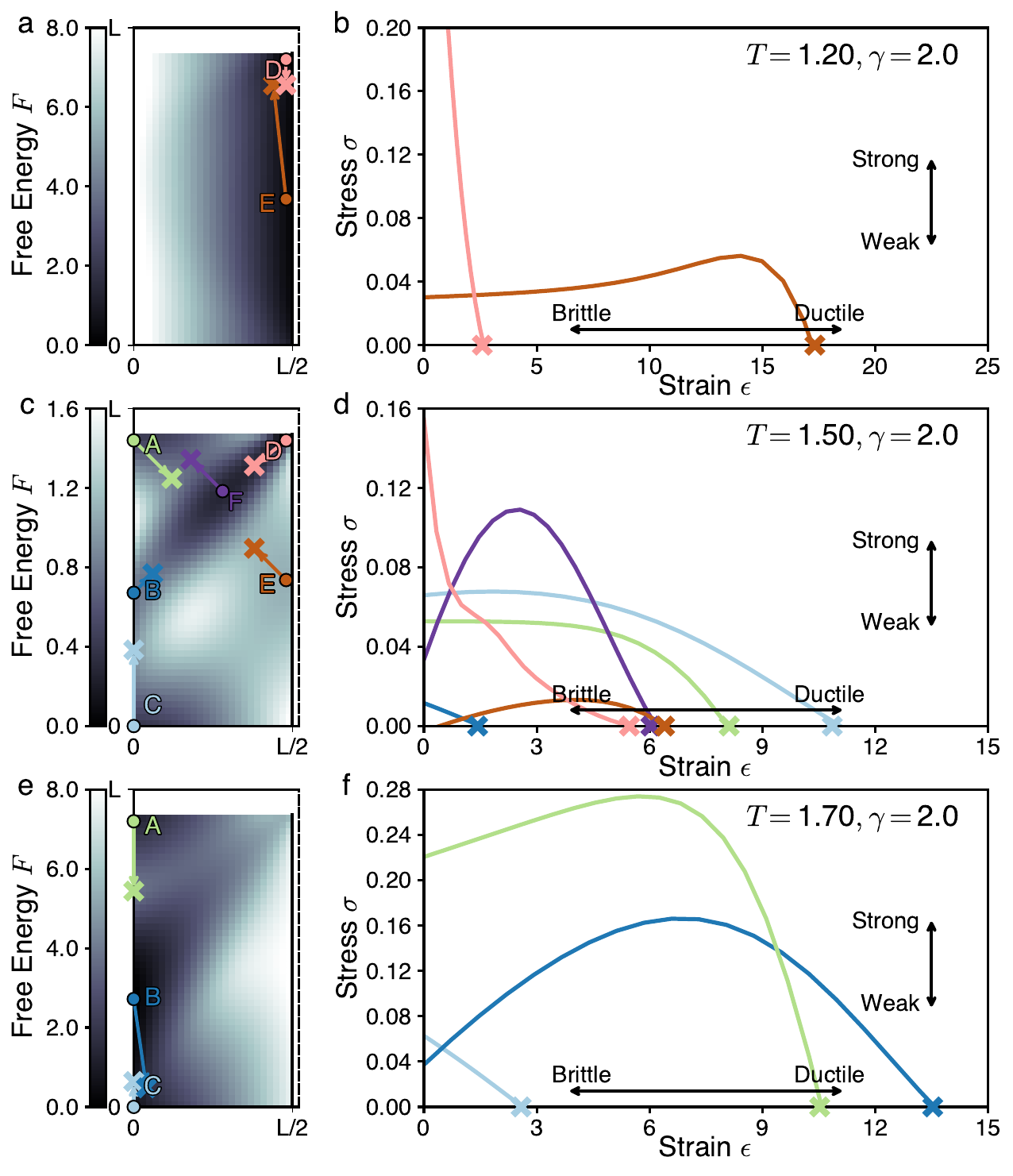}
\end{center}
\caption{
  Statistical physics approach quantifies stress--strain
  relationships in design problems. Plots show free
  energy landscapes and stress--strain curves at three different cost
  tolerances $T=1.20,1.50,1.70$ and constant bulkhead penalty $\gamma=2.0$.
  (a,c,e) Free energy landscape for the position of the left functional unit.
  The solid vertical line on the right denotes the position of the bulkhead. The
  dashed vertical line cuts off the domain of the second functional unit that
  has been integrated out. Note the different colormap scales at different $T$.
  Each colored circle indicates a local minimum that forms an architecture
  class, indexed with a unique letter $A$ through $F$ and a unique color (green,
  blue, purple etc.). The cross marks and lines connecting them to circles
  indicate the ultimate strain locations for each minima, as determined by
  condition \eqref{eqn:ult_strain}. (b,d,f) Stress-strain curves for each of the
  local minima at given $T$, with stress measured along the ultimate strain
  direction via spline interpolation of the free energy landscape. The cross
  marks indicate the ultimate strain $\epsilon^\text{ult}_k$ for each minimum.
  The maximum of each curve indicates the ultimate stress for each minimum
  $\sigma^\text{ult}_k$.}
\label{fig:strstr_panels}
\end{figure*}

Fig.~\ref{fig:strstr_panels} demonstrates the range of architecture classes and
their respective stress--strain curves that appear in the model subsystem. For
this subsystem, for each fixed $T$ in the range of $[0.5,2.0]$ we identify as
many as 6 qualitatively different architecture classes that we label $A$ through
$F$.  Figs.~\ref{fig:strstr_panels} and \ref{fig:minima_diagram} illustrate the
architecture classes at three representative values of $T={1.20,1.50,1.70}$.
$T=1.20$ corresponds to the low-cost regime. $T=1.70$ corresponds to the high
flexibility regime. $T=1.50$ corresponds to the regime in which there is a near
balance between cost and flexibility.

At each cost tolerance, for each architecture class we compute the
stress--strain response (Fig.~\ref{fig:strstr_panels}b,d,f). As described above,
for materials the stress--strain response can be measured via two principal
protocols. In the first, the material is deformed by a fixed strain
$\epsilon^\textsf{ext}$ and equilibrates at some corresponding stress
$\sigma(\epsilon^\textsf{ext})$. In the second, the material is affected by a
fixed stress $\sigma^\textsf{ext}$ and equilibrates at some corresponding strain
$\epsilon(\sigma^\textsf{ext})$. Graphically, this is equivalent to picking
first a point on the vertical $\sigma$ axis and finding the corresponding curve
point on the horizontal $\epsilon$ axis, or vice versa.

\subsubsection{Stress--Strain: Comparison with Materials}
For common materials,
the difference in protocols is not very noticeable since at low stress and
strain their relationship is linear, at larger deformations it is weakly
nonlinear but still monotonic, up until the breaking point at finite stress and
strain. However, this textbook materials science picture breaks down for
the design stress--strain relationship in our example design problem in three
important ways, at both low and high strain.

The differences between stress--strain relationships for designs and
materials can is facilitated by expanding the relationships at low strain as a
power series
\begin{equation}
  \sigma=\sigma_0 +Y\epsilon+O(\epsilon^2)\;.
\label{eqn:strstr_power}
\end{equation}

The first deviation from common material behavior is that for design stress the
constant term is nonzero $\sigma_0\neq 0$ for all of the curves in
Fig.~\ref{fig:strstr_panels}b,d,f.  Similar effects are common in manufactured
engineering components that exhibit residual internal stress, usually resulting
from plastic deformations in manufacturing, thermal expansion, boundary effects,
or phase change of materials.\cite{withers2007residual} The main implication of
residual stress is that applying small external design stress
$\sigma^\textsf{ext}<\sigma_0$ does not result in measurable design strain, as
opposed to the conventional linear response.

The second deviation is that the linear part of stress response can be both
positive ($Y>0$, as in Fig.~\ref{fig:strstr_panels}f, architecture classes
$A,B$, green and dark blue curves) and negative ($Y<0$, same figure,
architecture class $C$, light blue). A positive linear response means that the
architecture class can support at least small design stress above $\sigma_0$
level via a small deformation. A negative linear response means that the
ultimate stress $\sigma^\textsf{ult}=\sigma_0$ and is already reached for 
$\epsilon=0$, or no strain. For any higher, supercritical external stress, there
is no corresponding point $\epsilon(\sigma^\textsf{ext})$ and thus the
architecture class immediately ``breaks'', transitioning the design to a
different class.

Whereas the first two deviations from textbook materials response are observed
at low strain, the third one appears at high strain, right before the breaking
point. Many conventional materials (e.g.\ steel) break at finite stress.
However, some materials (e.g.\ fiber-reinforced brittle concrete
\cite{li1991micromechanical}) exhibit a different phenomenon known as ``tension
softening'',\cite{karihaloo1995fracture} whereby they support decreasing amounts
of stress as they are strained, and ultimately fail at zero stress.  We observe
this phenomenon in all of the architecture classes we find in the present model
system.

Together, these three deviations describe the unconventional pathways in which
architecture classes can break. Via strain: If the external subsystem coupling
provides a fixed design strain, the design stress remains positive and finite
for a wide strain range, ensuring that the chosen architecture class remains
viable. Via stress: Conversely, if the external subsystem coupling provides a
fixed stress, the architecture class only responds noticeably beyond a certain
stress threshold, but often responds with an abrupt architecture class change. 
Via stress and strain: a combination of external design stress and strain can
push the architecture class into the tension softening regime, making it
unviable. 

\subsubsection{Beyond Ultimate Stress}
\begin{figure}
\begin{center}
\includegraphics[width=0.5\textwidth]{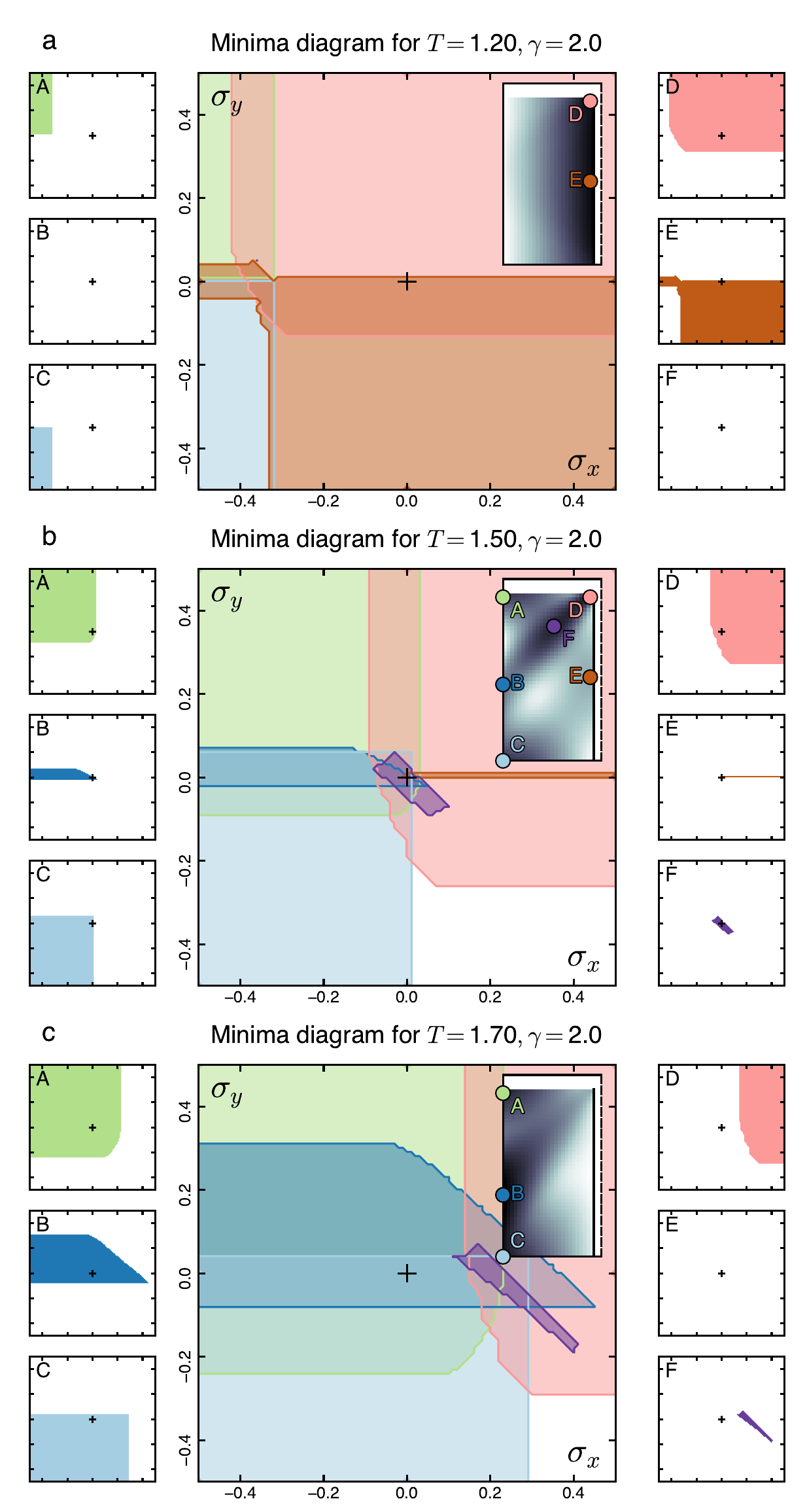}
\end{center}
\caption{
  The response to external stress gives viability limits on architecture
  classes and shows external stress can become viable under external stress.
Plots shows regions of existence of architecture classes in the
$(\sigma^\textsf{ext}_x,\sigma^\textsf{ext}_y)$ (external stress) plane. (a) $T=1.20$, (b)
  $T=1.50$, (c) $T=1.70$. For each panel, (center) Venn diagram of regions in
  the $(\sigma^\textsf{ext}_x,\sigma^\textsf{ext}_y)$ plane where each of 6
  architecture classes exists. Black cross indicates the origin of the plane
  $\vec{\sigma}^\textsf{ext}=0$. (inset) Free energy landscape with the
  architecture classes $A$ through $F$ labelled in color. (sides A--F) Regions in
  the $(\sigma^\textsf{ext}_x,\sigma^\textsf{ext}_y)$ plane where the
  corresponding individual minima exist, in the same stress plane as the central
  diagram.}
\label{fig:minima_diagram}
\end{figure}

Fig.~\ref{fig:strstr_panels} illustrates what happens to an architecture class
as it is strained up to its limit of viability by showing the design stress
along the direction of least ultimate strain. This is useful for assessing the
further viability of an architecture following the effects of an external design
strain. However, it can also be important to determine the effects on the number
of architecture classes under the influence of external design stress that can
push one or more classes beyond their limit of viability.

To assess robustness in this form,
Fig.~\ref{fig:minima_diagram} illustrates the effect of the external design
stress taking any values in the plane
$(\sigma^\textsf{ext}_x,\sigma^\textsf{ext}_y)$. In this plane, each
architecture class $A$ through $F$ has a viable domain. We find that the shapes
of these domains are complex, implying that the viability of an architecture
class is highly sensitive to both direction and magnitude of external stress. We
show the overlap of the domains of all six architecture classes in the center of
each panel in Fig.~\ref{fig:minima_diagram} in form of a computed Venn diagram.

We start analysis with the richest Venn diagram at the near-critical $T=1.50$ 
(Fig.~\ref{fig:minima_diagram}b). In that diagram, the classes $E,F$
(pink and purple) are viable in very narrow and specific ranges of external
stress $(\sigma^\textsf{ext}_x,\sigma^\textsf{ext}_y)$. Compared with the other
architectures, small amounts of uncertainty in external design stress would be
be sufficient to render architecture classes $E,F$ unviable. At the same time,
the architecture classes $A$ through $D$, in which the functional unit is
localized either in one of the three corners or the middle of one side of the
allowed domain, are viable given almost any amount of external design stress
outwards toward the domain boundaries, as well as moderate stress directed
inwards toward the middle of the domain. This form of analysis gives a more
detailed understanding variations in the robustness of architecture classes when
the effects of external design stress are not uniform in all directions.

From the depiction of the response to anisotropic stress in Fig.\
\ref{fig:minima_diagram} it can be seen that external stress can, indeed, push
some architectures beyond their viability limits while leaving others viable.
However, Fig.\ \ref{fig:minima_diagram} also shows that in situations where
there is low cost tolerance (i.e., a strong preference for low cost, $T=1.20$,
panel a) or high cost tolerance (i.e., a weak preference for low cost, $T=1.70$,
panel c), the external stress can make viable the architecture classes that would not be
viable without the external stress. Architectures $A$ and $C$ are examples of
this at low cost tolerance (panel a), whereas architectures $D$ and $F$ are
examples of this at high cost tolerance (panel c).

\subsubsection{Robustness and Design Pressure Changes}
\begin{figure*}
\begin{center}
\includegraphics[width=0.7\textwidth]{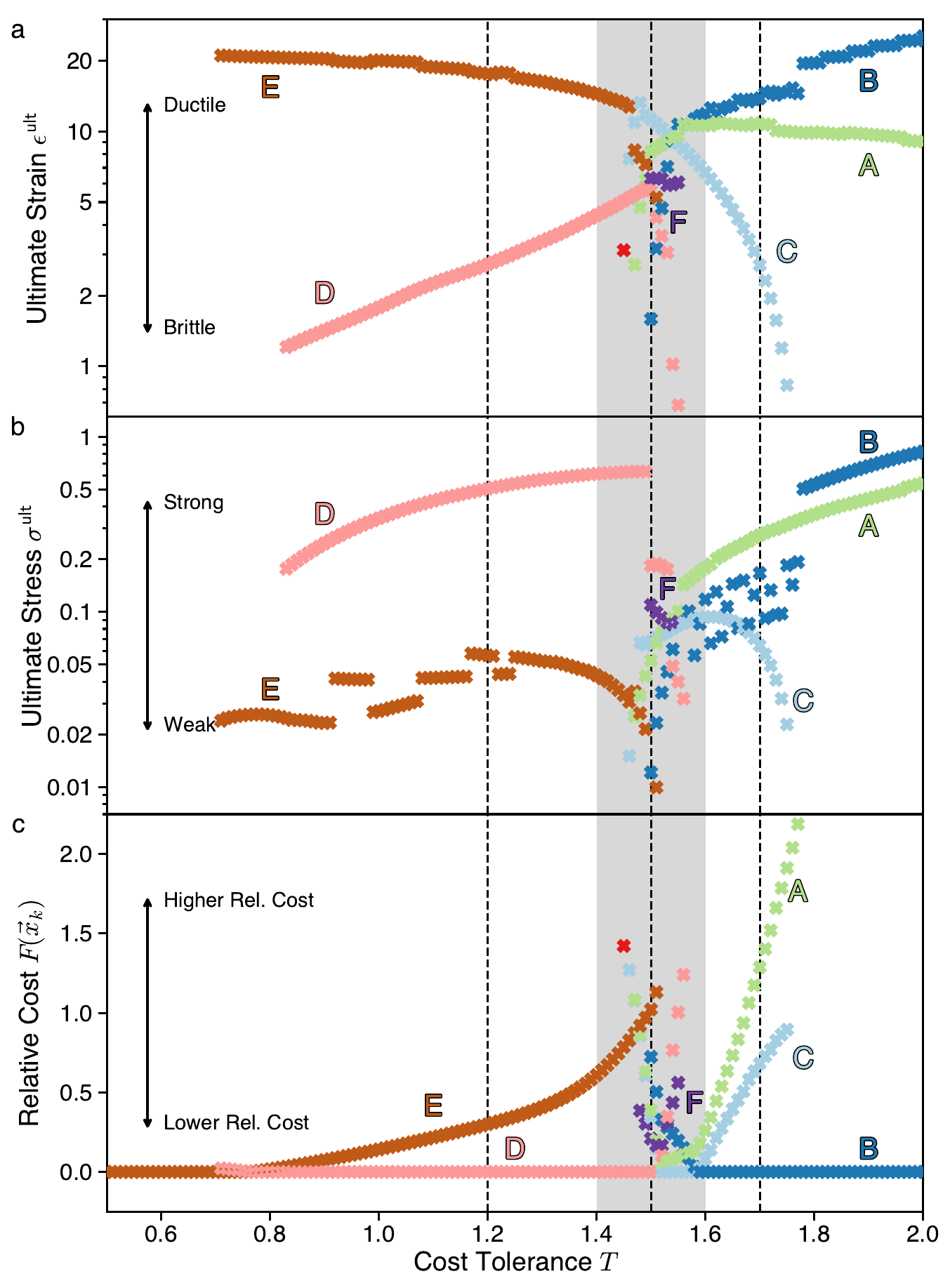}
\end{center}
\caption{Architecture classes can transition strong to weak, or from
brittle to ductile if design pressures change. Plots give robustness and
optimality measures for all architecture classes (local minima of Landau free
energy $F(\vec{x})$) existing at each value of cost
  tolerance $T$. The gray-shaded area on all three graphs indicates the region
  near $T_\textsf{crit}=1/\ln 2\approx 1.44$, which exhibits almost all of the
  architecture classes. The three dotted vertical lines indicate the values of
  $T$ for a detailed analysis is given in
  Figs.~\ref{fig:strstr_panels},~\ref{fig:minima_diagram}. Color and letter
  coding remain the same as in those Figures. (a) Ultimate strain values,
  corresponding to the \emph{brittle--ductile} robustness characterization of
  architecture classes. (b) Ultimate stress values, corresponding to the
  \emph{weak--strong} robustness characterization of architecture classes. (c)
  Landau free energy $F(\vec{x}_k)$ values at all local minima, normalized so
  that for any $T$ the lowest $F$ value is zero.  Vertical position of the
  points corresponds to \emph{higher-lower relative cost} characterization of
  architecture classes.}
\label{fig:ultimate_strstr}
\end{figure*}

Figs.~\ref{fig:strstr_panels}-\ref{fig:minima_diagram} showed a detailed
analysis of how particular architecture classes respond to external design
forcing at three representative values of cost tolerance $T$. To validate that
these choices of $T$ are representative, and to better understand how robustness
changes in response to changes in design pressure,
Fig.~\ref{fig:ultimate_strstr} aggregates the specific results shown in
Figs.~\ref{fig:strstr_panels}-\ref{fig:minima_diagram} with the results of
similar analysis over a fine grid of $T\in [0.5,2]$. As we scan the $T$ range,
the robustness of architecture classes experiences a clear shift in the region
of $T$ where the architectures reorganize from low-cost to high flexibility.
In the low-cost regime, architecture classes $D,E$ are viable, and
their properties are typified by the prior analysis at $T=1.20$.  In the
high-flexibility regime, architecture classes $A,B,C$ are viable, and their
properties typified by the prior analysis at $T=1.70$. At intermediate values of
$T$ where there is a reorganization between these regimes (shaded area in
Fig.~\ref{fig:ultimate_strstr}) we observe the highest diversity of viable
architectures at zero stress. However, we also observe sharp changes in the
robustness of architectures in response to external stress or strain. These change in 
robustness of the mesoscale design precisely in the near-critical $T$ region suggests a 
causal relationship: the shift between viable architecture classes is the primary 
mechanism to drive the large-scale phase transitions in the whole design space.

\subsubsection{Two-Factor Robustness Comparison of Architecture Classes}
\begin{figure*}
\begin{center}
\includegraphics[width=0.9\textwidth]{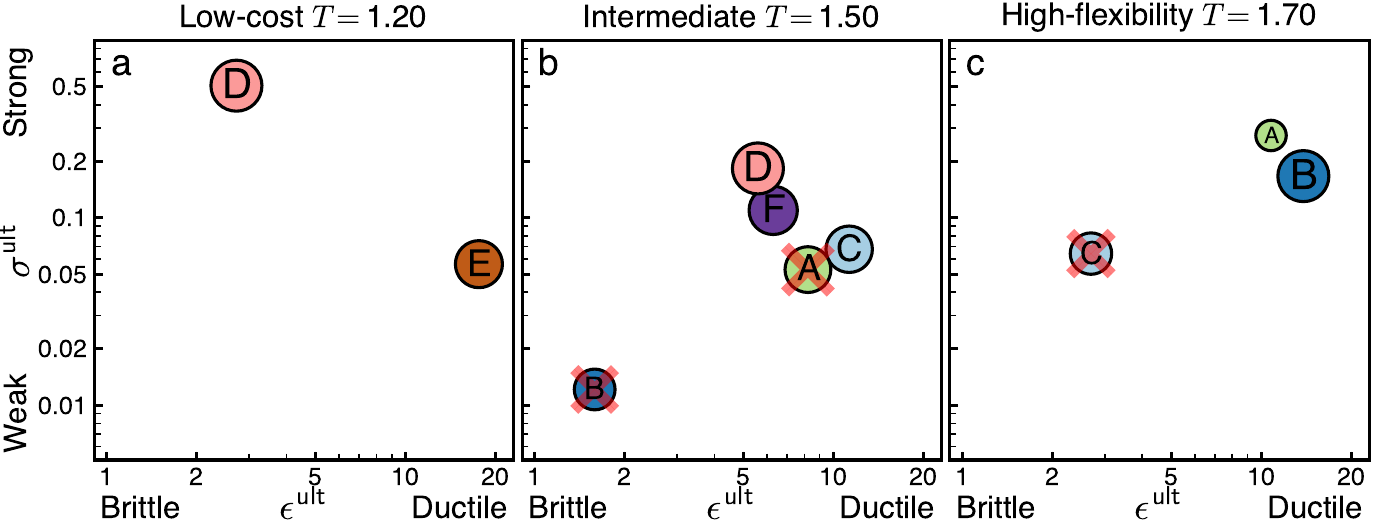}
\end{center}
\caption{Two-factor robustness ($R^2$) comparison for model-system
architecture classes facilitates the elimination of non-robust
(weak/brittle) architectures. The analysis of the model system implements the
scheme sketched in Fig.\ \ref{TwoAxes}, using the two
robustness factors of ultimate stress ($\epsilon^\textsf{ult}$, vertical axis)
and ultimate strain ($\sigma^\textsf{ult}$, horizontal axis). The three panels
depict the architectures and robustness relations among them at different
cost tolerances $T$ ($T=1.2$-a, $T=1.5$-b, $T=1.7$-c). Architectures are
represented by circles with size proportional to global design objectives. At
high cost tolerance (c), where design pressure favors flexibility in realizing
designs, the architecture C (also marked with a red $\times$) falls
in the shadow of both architectures A and B.  Between architectures A and B
there is a trade-off in robustness between strength and ductility. A similar
trade-off exists at low cost tolerance (a) between architectures D and E. At
intermediate tolerance (b) where there is a balance of concern between cost and
flexibility, more architectures are possible. Comparing their robustness,
the architecture A is eclipsed by C, the architecture B falls in the
shadow of all other architectures, and a trade-off exists between the three
architectures C, D, and F.
}
\label{fig:twofactor}
\end{figure*}

The analysis above gave detailed information about the structure of the design
space. Working from detailed information about the space, it is possible to
distill essential pieces of information that can inform design decisions.
Decisions involving comparing the robustness of architectures can be cast in the
form of $R^2$-plots. Whereas the Fig.~\ref{TwoAxes} diagrams were schematic
illustrations, applying the analysis framework illustrated in Fig.~\ref{TwoAxes}
to our model system yields the $R^2$-plots shown in
Fig.~\ref{fig:twofactor}.

Fig.~\ref{fig:twofactor} gives a proof-of-principle that two-factor comparisons
of architectures can be successfully computed in model systems. However, for a
given model system the comparisons themselves are interesting.  First, in
the model systems we study, we find no case in which a single architecture class
eclipses all others in robustness in both strength and ductility in a
$R^2$-plot. We find that, in the low-cost regime $T=1.20$
(Fig.~\ref{fig:twofactor}a), the trade-off is between the only two existing
architecture classes: architecture $D$ is stronger, but more brittle, whereas
architecture $E$ is weaker but more ductile.  In the regime where design
pressure favors designs that can be realized with more flexibility,
$T=1.70$ (Fig.~\ref{fig:twofactor}c), there is a strength/ductility trade-off
between architectures $A,B$, both of which eclipse architecture $C$. In the
intermediate regime, where design pressures for cost and flexibility are
nearly balanced, $T=1.50$ (Fig.~\ref{fig:twofactor}b), there is a three-way
strength/ductility trade-off between architectures $C,D,F$, but both of their
robustness factors are much smaller than at either higher or lower $T$. These
results indicate that the comparison of robustness between architectures is
complex. It depends both on the system being designed and on the design
pressures, and that even simple models can exhibit trade-offs.

\section{Conclusion}
In this paper we described a method to engineer systems that are robust to
variation in the design problem statement. We situated robustness in design as a
necessary complement to the robustness of a product under changes in the
operating environment (Fig.\ \ref{fig:robust_fragile}). We showed that the
robustness of a design is not a unitary concept, but instead it takes multiple
forms including being described on the axes brittle--ductile and weak--strong
(Figs.\ \ref{TwoAxes},\ref{fig:twofactor}). We found that these axes describe
system design in the same mathematical terms that are used to describe the
robustness of materials.  The approach developed in this paper sets the stage
for the further investigation of robust design.

First, to give a concrete demonstration of the application of our approach we
used an example model system drawn from arrangement problems in Naval
Engineering.\cite{Shields2017} In that model system we found that the design
architectures generically manifested forms of residual stress and tension
softening that are observed in materials systems.  Though both phenomena are
well-characterized in materials, they are phenomena that are associated
with special forms of preparation that introduce boundary effects
(residual stress) or by unconventional micromechanics in the material (tension
softening).\cite{li1991micromechanical} Our observation of these effects in a
model of system design raises the question of whether that observation is
specific to our model system, or whether system designs behave, in general, like
unconventional materials. The fact that the microscopic interactions in systems
design can easily have more varied forms than the micromechanics of materials
suggests that conventional behavior for systems may resemble unconventional
behavior in materials, but further investigation is required.

Second, though we needed to show the approach on a concrete model system, it
can be extended straightforwardly to other classes of design problems. We
believe that for a broad class of problems, the two-factor classification of
robustness ($R^2$-plot; Fig.\ \ref{TwoAxes}) will provide a straightforward means
for designers to compare the robustness of different design architecture
classes.
The comparison of classes in our example showed that the $R^2$ analysis
  provides a straightforward means to eliminate weak/brittle architecture
  classes. We expect this to be useful in complex system design where it has
  been argued that eliminating bad choices is more important than selecting good
  ones.\cite{singersbd}
  The nontrivial trade-offs in $R^2$ we found here suggest that
  a similar multifaceted characterization will be useful in broader
  classes of complex systems design.

\section*{Acknowledgements}
We thank two anonymous reviewers for suggestions that improved the
readability of the manuscript. This work was supported by the U.S.\ Office of
Naval Research Grant Nos.\ N00014-17-1-2491 and N00014-15-1-275. This research
was supported in part through computational resources and services provided by
Advanced Research Computing at the University of Michigan, Ann Arbor. GvA
acknowledges the support of the Natural Sciences and Engineering Research
Council of Canada (NSERC).

\section*{Author contributions statement}
AAK and AK developed and analyzed mathematical models. AAK and GvA wrote the
paper in consultation with AK and DJS. DJS provided the naval engineering
context and expertise. GvA initiated and supervised the research.

\section*{Additional information}
We declare we have no competing interests.


\begin{thebibliography}{40}%
	\makeatletter
	\providecommand \@ifxundefined [1]{%
		\@ifx{#1\undefined}
	}%
	\providecommand \@ifnum [1]{%
		\ifnum #1\expandafter \@firstoftwo
		\else \expandafter \@secondoftwo
		\fi
	}%
	\providecommand \@ifx [1]{%
		\ifx #1\expandafter \@firstoftwo
		\else \expandafter \@secondoftwo
		\fi
	}%
	\providecommand \natexlab [1]{#1}%
	\providecommand \enquote  [1]{``#1''}%
	\providecommand \bibnamefont  [1]{#1}%
	\providecommand \bibfnamefont [1]{#1}%
	\providecommand \citenamefont [1]{#1}%
	\providecommand \href@noop [0]{\@secondoftwo}%
	\providecommand \href [0]{\begingroup \@sanitize@url \@href}%
	\providecommand \@href[1]{\@@startlink{#1}\@@href}%
	\providecommand \@@href[1]{\endgroup#1\@@endlink}%
	\providecommand \@sanitize@url [0]{\catcode `\\12\catcode `\$12\catcode
		`\&12\catcode `\#12\catcode `\^12\catcode `\_12\catcode `\%12\relax}%
	\providecommand \@@startlink[1]{}%
	\providecommand \@@endlink[0]{}%
	\providecommand \url  [0]{\begingroup\@sanitize@url \@url }%
	\providecommand \@url [1]{\endgroup\@href {#1}{\urlprefix }}%
	\providecommand \urlprefix  [0]{URL }%
	\providecommand \Eprint [0]{\href }%
	\providecommand \doibase [0]{http://dx.doi.org/}%
	\providecommand \selectlanguage [0]{\@gobble}%
	\providecommand \bibinfo  [0]{\@secondoftwo}%
	\providecommand \bibfield  [0]{\@secondoftwo}%
	\providecommand \translation [1]{[#1]}%
	\providecommand \BibitemOpen [0]{}%
	\providecommand \bibitemStop [0]{}%
	\providecommand \bibitemNoStop [0]{.\EOS\space}%
	\providecommand \EOS [0]{\spacefactor3000\relax}%
	\providecommand \BibitemShut  [1]{\csname bibitem#1\endcsname}%
	\let\auto@bib@innerbib\@empty
	\bibitem [{\citenamefont {Taguchi}(1986)}]{taguchi1986introduction}%
	\BibitemOpen
	\bibfield  {author} {\bibinfo {author} {\bibfnamefont {G.}~\bibnamefont
			{Taguchi}},\ }\href@noop {} {\emph {\bibinfo {title} {Introduction 
			to quality
				engineering: designing quality into products and processes}}}\ 
				(\bibinfo
	{publisher} {Asian Productivity Association},\ \bibinfo {address} {Tokyo},\
	\bibinfo {year} {1986})\BibitemShut {NoStop}%
	\bibitem [{\citenamefont {Phadke}(1995)}]{phadke1995quality}%
	\BibitemOpen
	\bibfield  {author} {\bibinfo {author} {\bibfnamefont {M.~S.}\ \bibnamefont
			{Phadke}},\ }\href@noop {} {\emph {\bibinfo {title} {Quality 
			engineering
				using robust design}}}\ (\bibinfo  {publisher} {Prentice Hall 
				PTR},\ \bibinfo
	{year} {1995})\BibitemShut {NoStop}%
	\bibitem [{\citenamefont {Boehm}(1988)}]{boehm1988software}%
	\BibitemOpen
	\bibfield  {author} {\bibinfo {author} {\bibfnamefont {B.~W.}\ \bibnamefont
			{Boehm}},\ }\href {\doibase 10.1109/32.6191} {\bibfield  {journal} 
			{\bibinfo
			{journal} {Journal of Parametrics}\ }\textbf {\bibinfo {volume} 
			{8}},\
		\bibinfo {pages} {32} (\bibinfo {year} {1988})}\BibitemShut {NoStop}%
	\bibitem [{\citenamefont {Love}\ \emph {et~al.}(1999)\citenamefont {Love},
		\citenamefont {Mandal},\ and\ \citenamefont {Li}}]{love1999determining}%
	\BibitemOpen
	\bibfield  {author} {\bibinfo {author} {\bibfnamefont {P.~E.}\ \bibnamefont
			{Love}}, \bibinfo {author} {\bibfnamefont {P.}~\bibnamefont 
			{Mandal}}, \ and\
		\bibinfo {author} {\bibfnamefont {H.}~\bibnamefont {Li}},\ }\href 
		{\doibase
		10.1080/014461999371420} {\bibfield  {journal} {\bibinfo  {journal}
			{Construction Management \& Economics}\ }\textbf {\bibinfo {volume} 
			{17}},\
		\bibinfo {pages} {505} (\bibinfo {year} {1999})}\BibitemShut {NoStop}%
	\bibitem [{\citenamefont {Manyika}\ \emph {et~al.}(2012)\citenamefont
		{Manyika}, \citenamefont {Sinclair}, \citenamefont {Dobbs}, 
		\citenamefont
		{Strube}, \citenamefont {Rassey}, \citenamefont {Mischke}, \citenamefont
		{Remes}, \citenamefont {Roxburgh}, \citenamefont {George}, \citenamefont
		{O'Halloran},\ and\ \citenamefont {Ramaswamy}}]{mckinseymfg}%
	\BibitemOpen
	\bibfield  {author} {\bibinfo {author} {\bibfnamefont {J.}~\bibnamefont
			{Manyika}}, \bibinfo {author} {\bibfnamefont {J.}~\bibnamefont 
			{Sinclair}},
		\bibinfo {author} {\bibfnamefont {R.}~\bibnamefont {Dobbs}}, \bibinfo
		{author} {\bibfnamefont {G.}~\bibnamefont {Strube}}, \bibinfo {author}
		{\bibfnamefont {L.}~\bibnamefont {Rassey}}, \bibinfo {author} 
		{\bibfnamefont
			{J.}~\bibnamefont {Mischke}}, \bibinfo {author} {\bibfnamefont
			{J.}~\bibnamefont {Remes}}, \bibinfo {author} {\bibfnamefont
			{C.}~\bibnamefont {Roxburgh}}, \bibinfo {author} {\bibfnamefont
			{K.}~\bibnamefont {George}}, \bibinfo {author} {\bibfnamefont
			{D.}~\bibnamefont {O'Halloran}}, \ and\ \bibinfo {author} 
			{\bibfnamefont
			{S.}~\bibnamefont {Ramaswamy}},\ }\href
	{http://www.mckinsey.com/business-functions/operations/our-insights/the-future-of-manufacturing}
	{\emph {\bibinfo {title} {Manufacturing the future: The next era of global
				growth and innovation}}}\ (\bibinfo  {publisher} {McKinsey 
				Global
		Institute},\ \bibinfo {year} {2012})\BibitemShut {NoStop}%
	\bibitem [{\citenamefont {Shields}\ \emph {et~al.}(2017)\citenamefont
		{Shields}, \citenamefont {Rigterink},\ and\ \citenamefont
		{Singer}}]{Shields2017}%
	\BibitemOpen
	\bibfield  {author} {\bibinfo {author} {\bibfnamefont {C.~P.}\ \bibnamefont
			{Shields}}, \bibinfo {author} {\bibfnamefont {D.~T.}\ \bibnamefont
			{Rigterink}}, \ and\ \bibinfo {author} {\bibfnamefont {D.~J.}\ 
			\bibnamefont
			{Singer}},\ }\href {\doibase 10.1016/j.oceaneng.2017.02.037} 
			{\bibfield
		{journal} {\bibinfo  {journal} {Ocean Eng.}\ }\textbf {\bibinfo {volume}
			{135}},\ \bibinfo {pages} {236 } (\bibinfo {year} 
			{2017})}\BibitemShut
	{NoStop}%
	\bibitem [{\citenamefont {Dieter}\ and\ \citenamefont
		{Bacon}(1986)}]{DieterBacon}%
	\BibitemOpen
	\bibfield  {author} {\bibinfo {author} {\bibfnamefont {G.~E.}\ \bibnamefont
			{Dieter}}\ and\ \bibinfo {author} {\bibfnamefont {D.~J.}\ 
			\bibnamefont
			{Bacon}},\ }\href@noop {} {\emph {\bibinfo {title} {Mechanical
				Metallurgy}}},\ \bibinfo {edition} {3rd}\ ed.\ (\bibinfo  
				{publisher}
	{McGraw-Hill New York},\ \bibinfo {year} {1986})\BibitemShut {NoStop}%
	\bibitem [{\citenamefont {Li}\ \emph {et~al.}(2012)\citenamefont {Li},
		\citenamefont {Bashan}, \citenamefont {Buldyrev}, \citenamefont 
		{Stanley},\
		and\ \citenamefont {Havlin}}]{Stanley2012failures}%
	\BibitemOpen
	\bibfield  {author} {\bibinfo {author} {\bibfnamefont {W.}~\bibnamefont
			{Li}}, \bibinfo {author} {\bibfnamefont {A.}~\bibnamefont 
			{Bashan}}, \bibinfo
		{author} {\bibfnamefont {S.~V.}\ \bibnamefont {Buldyrev}}, \bibinfo 
		{author}
		{\bibfnamefont {H.~E.}\ \bibnamefont {Stanley}}, \ and\ \bibinfo 
		{author}
		{\bibfnamefont {S.}~\bibnamefont {Havlin}},\ }\href {\doibase
		10.1103/PhysRevLett.108.228702} {\bibfield  {journal} {\bibinfo  
		{journal}
			{Phys. Rev. Lett.}\ }\textbf {\bibinfo {volume} {108}},\ \bibinfo 
			{pages}
		{228702} (\bibinfo {year} {2012})}\BibitemShut {NoStop}%
	\bibitem [{\citenamefont {Buldyrev}\ \emph {et~al.}(2010)\citenamefont
		{Buldyrev}, \citenamefont {Parshani}, \citenamefont {Paul}, 
		\citenamefont
		{Stanley},\ and\ \citenamefont {Havlin}}]{buldyrev2010catastrophic}%
	\BibitemOpen
	\bibfield  {author} {\bibinfo {author} {\bibfnamefont {S.~V.}\ \bibnamefont
			{Buldyrev}}, \bibinfo {author} {\bibfnamefont {R.}~\bibnamefont 
			{Parshani}},
		\bibinfo {author} {\bibfnamefont {G.}~\bibnamefont {Paul}}, \bibinfo 
		{author}
		{\bibfnamefont {H.~E.}\ \bibnamefont {Stanley}}, \ and\ \bibinfo 
		{author}
		{\bibfnamefont {S.}~\bibnamefont {Havlin}},\ }\href {\doibase
		10.1038/nature08932} {\bibfield  {journal} {\bibinfo  {journal} 
		{Nature}\
		}\textbf {\bibinfo {volume} {464}},\ \bibinfo {pages} {1025} (\bibinfo 
		{year}
		{2010})}\BibitemShut {NoStop}%
	\bibitem [{\citenamefont {Brummitt}\ \emph {et~al.}(2012)\citenamefont
		{Brummitt}, \citenamefont {D{\textquoteright}Souza},\ and\ \citenamefont
		{Leicht}}]{Brummitt2012}%
	\BibitemOpen
	\bibfield  {author} {\bibinfo {author} {\bibfnamefont {C.~D.}\ \bibnamefont
			{Brummitt}}, \bibinfo {author} {\bibfnamefont {R.~M.}\ \bibnamefont
			{D{\textquoteright}Souza}}, \ and\ \bibinfo {author} {\bibfnamefont 
			{E.~A.}\
			\bibnamefont {Leicht}},\ }\href {\doibase 10.1073/pnas.1110586109} 
			{\bibfield
		{journal} {\bibinfo  {journal} {Proc. Natl. Acad. Sci. U.S.A.}\ }\textbf
		{\bibinfo {volume} {109}},\ \bibinfo {pages} {E680} (\bibinfo {year}
		{2012})}\BibitemShut {NoStop}%
	\bibitem [{\citenamefont {Klishin}\ \emph {et~al.}(2018)\citenamefont
		{Klishin}, \citenamefont {Shields}, \citenamefont {Singer},\ and\
		\citenamefont {van Anders}}]{systemphys}%
	\BibitemOpen
	\bibfield  {author} {\bibinfo {author} {\bibfnamefont {A.~A.}\ \bibnamefont
			{Klishin}}, \bibinfo {author} {\bibfnamefont {C.~P.}\ \bibnamefont
			{Shields}}, \bibinfo {author} {\bibfnamefont {D.~J.}\ \bibnamefont 
			{Singer}},
		\ and\ \bibinfo {author} {\bibfnamefont {G.}~\bibnamefont {van 
		Anders}},\
	}\href {\doibase 10.1088/1367-2630/aae72a} {\bibfield  {journal} {\bibinfo
			{journal} {New J. Phys.}\ }\textbf {\bibinfo {volume} {20}},\ 
			\bibinfo
		{pages} {103038} (\bibinfo {year} {2018})},\ \Eprint
	{http://arxiv.org/abs/1709.03388} {arXiv:1709.03388 [physics.soc-ph]}
	\BibitemShut {NoStop}%
	\bibitem [{\citenamefont {Evans}(1959)}]{evans1959}%
	\BibitemOpen
	\bibfield  {author} {\bibinfo {author} {\bibfnamefont {J.~H.}\ \bibnamefont
			{Evans}},\ }\href {\doibase 10.1111/j.1559-3584.1959.tb01836.x} 
			{\bibfield
		{journal} {\bibinfo  {journal} {Naval Engineers Journal}\ }\textbf 
		{\bibinfo
			{volume} {71}},\ \bibinfo {pages} {671} (\bibinfo {year} 
			{1959})}\BibitemShut
	{NoStop}%
	\bibitem [{\citenamefont {Papalambros}\ and\ \citenamefont
		{Wilde}(2000)}]{papalambros2000principles}%
	\BibitemOpen
	\bibfield  {author} {\bibinfo {author} {\bibfnamefont {P.~Y.}\ \bibnamefont
			{Papalambros}}\ and\ \bibinfo {author} {\bibfnamefont {D.~J.}\ 
			\bibnamefont
			{Wilde}},\ }\href@noop {} {\emph {\bibinfo {title} {Principles of 
			optimal
				design: modeling and computation}}}\ (\bibinfo  {publisher} 
				{Cambridge
		University Press},\ \bibinfo {year} {2000})\BibitemShut {NoStop}%
	\bibitem [{\citenamefont {Chalfant}(2015)}]{chalfant2015}%
	\BibitemOpen
	\bibfield  {author} {\bibinfo {author} {\bibfnamefont {J.}~\bibnamefont
			{Chalfant}},\ }\href {\doibase 10.1109/JPROC.2015.2459672} 
			{\bibfield
		{journal} {\bibinfo  {journal} {Proceedings of the IEEE}\ }\textbf 
		{\bibinfo
			{volume} {103}},\ \bibinfo {pages} {2252} (\bibinfo {year}
		{2015})}\BibitemShut {NoStop}%
	\bibitem [{\citenamefont {Andrews}(2012)}]{Andrews2012}%
	\BibitemOpen
	\bibfield  {author} {\bibinfo {author} {\bibfnamefont {D.~J.}\ \bibnamefont
			{Andrews}},\ }\href {\doibase 10.1098/rspa.2011.0590} {\bibfield  
			{journal}
		{\bibinfo  {journal} {Proc. Roy. Soc. London A}\ }\textbf {\bibinfo 
		{volume}
			{468}},\ \bibinfo {pages} {891} (\bibinfo {year} 
			{2012})}\BibitemShut
	{NoStop}%
	\bibitem [{\citenamefont {Liker}(2004)}]{2004toyota}%
	\BibitemOpen
	\bibfield  {author} {\bibinfo {author} {\bibfnamefont {J.~K.}\ \bibnamefont
			{Liker}},\ }\href@noop {} {\emph {\bibinfo {title} {The Toyota 
			Way}}}\
	(\bibinfo  {publisher} {McGraw-Hill},\ \bibinfo {address} {New York},\
	\bibinfo {year} {2004})\BibitemShut {NoStop}%
	\bibitem [{\citenamefont {Morgan}\ and\ \citenamefont
		{Liker}(2006)}]{morgan2006toyota}%
	\BibitemOpen
	\bibfield  {author} {\bibinfo {author} {\bibfnamefont {J.~M.}\ \bibnamefont
			{Morgan}}\ and\ \bibinfo {author} {\bibfnamefont {J.~K.}\ 
			\bibnamefont
			{Liker}},\ }\href@noop {} {\emph {\bibinfo {title} {The Toyota 
			product
				development system}}},\ Vol.\ \bibinfo {volume} {13533}\ 
				(\bibinfo
	{publisher} {Productivity Press},\ \bibinfo {address} {New York},\ \bibinfo
	{year} {2006})\BibitemShut {NoStop}%
	\bibitem [{\citenamefont {Bernstein}(1998)}]{bernstein1998}%
	\BibitemOpen
	\bibfield  {author} {\bibinfo {author} {\bibfnamefont {J.~I.}\ \bibnamefont
			{Bernstein}},\ }\emph {\bibinfo {title} {Design methods in the 
			aerospace
			industry: looking for evidence of set-based practices}},\ 
			\href@noop {}
	{Ph.D. thesis},\ \bibinfo  {school} {Massachusetts Institute of Technology}
	(\bibinfo {year} {1998})\BibitemShut {NoStop}%
	\bibitem [{\citenamefont {Singer}\ \emph {et~al.}(2009)\citenamefont 
	{Singer},
		\citenamefont {Doerry},\ and\ \citenamefont {Buckley}}]{singersbd}%
	\BibitemOpen
	\bibfield  {author} {\bibinfo {author} {\bibfnamefont {D.~J.}\ \bibnamefont
			{Singer}}, \bibinfo {author} {\bibfnamefont {N.}~\bibnamefont 
			{Doerry}}, \
		and\ \bibinfo {author} {\bibfnamefont {M.~E.}\ \bibnamefont {Buckley}},\
	}\href {\doibase 10.1111/j.1559-3584.2009.00226.x} {\bibfield  {journal}
		{\bibinfo  {journal} {Naval Engineers Journal}\ }\textbf {\bibinfo 
		{volume}
			{121}},\ \bibinfo {pages} {31} (\bibinfo {year} {2009})}\BibitemShut
	{NoStop}%
	\bibitem [{\citenamefont {Ben-Tal}\ and\ \citenamefont
		{Nemirovski}(2000)}]{bental2000robust}%
	\BibitemOpen
	\bibfield  {author} {\bibinfo {author} {\bibfnamefont {A.}~\bibnamefont
			{Ben-Tal}}\ and\ \bibinfo {author} {\bibfnamefont {A.}~\bibnamefont
			{Nemirovski}},\ }\href {\doibase 10.1007/s101070000163} {\bibfield  
			{journal}
		{\bibinfo  {journal} {Mathematical Programming}\ }\textbf {\bibinfo 
		{volume}
			{88}},\ \bibinfo {pages} {411} (\bibinfo {year} {2000})}\BibitemShut
	{NoStop}%
	\bibitem [{\citenamefont {Bertsimas}\ \emph {et~al.}(2011)\citenamefont
		{Bertsimas}, \citenamefont {Brown},\ and\ \citenamefont
		{Caramanis}}]{bertsimas2011theory}%
	\BibitemOpen
	\bibfield  {author} {\bibinfo {author} {\bibfnamefont {D.}~\bibnamefont
			{Bertsimas}}, \bibinfo {author} {\bibfnamefont {D.~B.}\ \bibnamefont
			{Brown}}, \ and\ \bibinfo {author} {\bibfnamefont {C.}~\bibnamefont
			{Caramanis}},\ }\href {\doibase 10.1137/080734510} {\bibfield  
			{journal}
		{\bibinfo  {journal} {SIAM Rev.}\ }\textbf {\bibinfo {volume} {53}},\
		\bibinfo {pages} {464} (\bibinfo {year} {2011})}\BibitemShut {NoStop}%
	\bibitem [{\citenamefont {Gibbs}(1902)}]{gibbs}%
	\BibitemOpen
	\bibfield  {author} {\bibinfo {author} {\bibfnamefont {J.~W.}\ \bibnamefont
			{Gibbs}},\ }\href@noop {} {\emph {\bibinfo {title} {Elementary 
			Principles in
				Statistical Mechanics}}}\ (\bibinfo  {publisher} {Charles 
				Scribner's Sons},\
	\bibinfo {address} {New York},\ \bibinfo {year} {1902})\BibitemShut 
	{NoStop}%
	\bibitem [{\citenamefont {van Anders}\ \emph {et~al.}(2014)\citenamefont 
	{van
			Anders}, \citenamefont {Klotsa}, \citenamefont {Ahmed}, 
			\citenamefont
		{Engel},\ and\ \citenamefont {Glotzer}}]{entint}%
	\BibitemOpen
	\bibfield  {author} {\bibinfo {author} {\bibfnamefont {G.}~\bibnamefont {van
				Anders}}, \bibinfo {author} {\bibfnamefont {D.}~\bibnamefont 
				{Klotsa}},
		\bibinfo {author} {\bibfnamefont {N.~K.}\ \bibnamefont {Ahmed}}, 
		\bibinfo
		{author} {\bibfnamefont {M.}~\bibnamefont {Engel}}, \ and\ \bibinfo 
		{author}
		{\bibfnamefont {S.~C.}\ \bibnamefont {Glotzer}},\ }\href {\doibase
		10.1073/pnas.1418159111} {\bibfield  {journal} {\bibinfo  {journal} 
		{Proc.
				Natl. Acad. Sci. U.S.A.}\ }\textbf {\bibinfo {volume} {111}},\ 
				\bibinfo
		{pages} {E4812} (\bibinfo {year} {2014})},\ \Eprint
	{http://arxiv.org/abs/1309.1187} {arXiv:1309.1187 [cond-mat.soft]}
	\BibitemShut {NoStop}%
	\bibitem [{\citenamefont {Frenkel}(2015)}]{ordviaent}%
	\BibitemOpen
	\bibfield  {author} {\bibinfo {author} {\bibfnamefont {D.}~\bibnamefont
			{Frenkel}},\ }\href {\doibase 10.1038/nmat4178} {\bibfield  
			{journal}
		{\bibinfo  {journal} {Nat. Mater.}\ }\textbf {\bibinfo {volume} {14}},\
		\bibinfo {pages} {9} (\bibinfo {year} {2015})}\BibitemShut {NoStop}%
	\bibitem [{\citenamefont {Geng}\ \emph {et~al.}(2019)\citenamefont {Geng},
		\citenamefont {van Anders}, \citenamefont {Dodd}, \citenamefont
		{Dshemuchadse},\ and\ \citenamefont {Glotzer}}]{engent}%
	\BibitemOpen
	\bibfield  {author} {\bibinfo {author} {\bibfnamefont {Y.}~\bibnamefont
			{Geng}}, \bibinfo {author} {\bibfnamefont {G.}~\bibnamefont {van 
			Anders}},
		\bibinfo {author} {\bibfnamefont {P.~M.}\ \bibnamefont {Dodd}}, \bibinfo
		{author} {\bibfnamefont {J.}~\bibnamefont {Dshemuchadse}}, \ and\ 
		\bibinfo
		{author} {\bibfnamefont {S.~C.}\ \bibnamefont {Glotzer}},\ }\href 
		{\doibase
		10.1126/sciadv.aaw0514} {\bibfield  {journal} {\bibinfo  {journal} 
		{Science
				Advances}\ }\textbf {\bibinfo {volume} {5}},\ \bibinfo {pages} 
				{eaaw0514}
		(\bibinfo {year} {2019})}\BibitemShut {NoStop}%
	\bibitem [{\citenamefont {Harper}\ \emph {et~al.}(2019)\citenamefont 
	{Harper},
		\citenamefont {van Anders},\ and\ \citenamefont {Glotzer}}]{entchem}%
	\BibitemOpen
	\bibfield  {author} {\bibinfo {author} {\bibfnamefont {E.~S.}\ \bibnamefont
			{Harper}}, \bibinfo {author} {\bibfnamefont {G.}~\bibnamefont {van 
			Anders}},
		\ and\ \bibinfo {author} {\bibfnamefont {S.~C.}\ \bibnamefont 
		{Glotzer}},\
	}\href {\doibase 10.1073/pnas.1822092116} {\bibfield  {journal} {\bibinfo
			{journal} {Proc. Natl. Acad. Sci. U.S.A.}\ }\textbf {\bibinfo 
			{volume}
			{116}},\ \bibinfo {pages} {16703} (\bibinfo {year} 
			{2019})}\BibitemShut
	{NoStop}%
	\bibitem [{\citenamefont {Vicsek}\ and\ \citenamefont
		{Zafeiris}(2012)}]{vicsek2012collective}%
	\BibitemOpen
	\bibfield  {author} {\bibinfo {author} {\bibfnamefont {T.}~\bibnamefont
			{Vicsek}}\ and\ \bibinfo {author} {\bibfnamefont {A.}~\bibnamefont
			{Zafeiris}},\ }\href {\doibase 10.1016/j.physrep.2012.03.004} 
			{\bibfield
		{journal} {\bibinfo  {journal} {Physics Reports}\ }\textbf {\bibinfo 
		{volume}
			{517}},\ \bibinfo {pages} {71} (\bibinfo {year} {2012})}\BibitemShut
	{NoStop}%
	\bibitem [{\citenamefont {Silverberg}\ \emph {et~al.}(2013)\citenamefont
		{Silverberg}, \citenamefont {Bierbaum}, \citenamefont {Sethna},\ and\
		\citenamefont {Cohen}}]{silverberg2013moshpit}%
	\BibitemOpen
	\bibfield  {author} {\bibinfo {author} {\bibfnamefont {J.~L.}\ \bibnamefont
			{Silverberg}}, \bibinfo {author} {\bibfnamefont {M.}~\bibnamefont
			{Bierbaum}}, \bibinfo {author} {\bibfnamefont {J.~P.}\ \bibnamefont
			{Sethna}}, \ and\ \bibinfo {author} {\bibfnamefont {I.}~\bibnamefont
			{Cohen}},\ }\href {\doibase 10.1103/PhysRevLett.110.228701} 
			{\bibfield
		{journal} {\bibinfo  {journal} {Phys. Rev. Lett.}\ }\textbf {\bibinfo
			{volume} {110}},\ \bibinfo {pages} {228701} (\bibinfo {year}
		{2013})}\BibitemShut {NoStop}%
	\bibitem [{\citenamefont {Chowdhury}\ \emph {et~al.}(2000)\citenamefont
		{Chowdhury}, \citenamefont {Santen},\ and\ \citenamefont
		{Schadschneider}}]{chowdhury2000statistical}%
	\BibitemOpen
	\bibfield  {author} {\bibinfo {author} {\bibfnamefont {D.}~\bibnamefont
			{Chowdhury}}, \bibinfo {author} {\bibfnamefont {L.}~\bibnamefont 
			{Santen}}, \
		and\ \bibinfo {author} {\bibfnamefont {A.}~\bibnamefont 
		{Schadschneider}},\
	}\href {\doibase 10.1016/S0370-1573(99)00117-9} {\bibfield  {journal}
		{\bibinfo  {journal} {Physics Reports}\ }\textbf {\bibinfo {volume} 
		{329}},\
		\bibinfo {pages} {199} (\bibinfo {year} {2000})}\BibitemShut {NoStop}%
	\bibitem [{\citenamefont {Noble}\ \emph {et~al.}(2018)\citenamefont {Noble},
		\citenamefont {Rosenstock}, \citenamefont {Brown}, \citenamefont 
		{Machta},\
		and\ \citenamefont {Hastings}}]{noble2018spatial}%
	\BibitemOpen
	\bibfield  {author} {\bibinfo {author} {\bibfnamefont {A.~E.}\ \bibnamefont
			{Noble}}, \bibinfo {author} {\bibfnamefont {T.~S.}\ \bibnamefont
			{Rosenstock}}, \bibinfo {author} {\bibfnamefont {P.~H.}\ 
			\bibnamefont
			{Brown}}, \bibinfo {author} {\bibfnamefont {J.}~\bibnamefont 
			{Machta}}, \
		and\ \bibinfo {author} {\bibfnamefont {A.}~\bibnamefont {Hastings}},\ 
		}\href
	{\doibase 10.1073/pnas.1618887115} {\bibfield  {journal} {\bibinfo  
	{journal}
			{Proceedings of the National Academy of Sciences}\ }\textbf 
			{\bibinfo
			{volume} {115}},\ \bibinfo {pages} {1825} (\bibinfo {year}
		{2018})}\BibitemShut {NoStop}%
	\bibitem [{\citenamefont {Daniels}\ \emph {et~al.}(2017)\citenamefont
		{Daniels}, \citenamefont {Krakauer},\ and\ \citenamefont
		{Flack}}]{flack2017critical}%
	\BibitemOpen
	\bibfield  {author} {\bibinfo {author} {\bibfnamefont {B.~C.}\ \bibnamefont
			{Daniels}}, \bibinfo {author} {\bibfnamefont {D.~C.}\ \bibnamefont
			{Krakauer}}, \ and\ \bibinfo {author} {\bibfnamefont {J.~C.}\ 
			\bibnamefont
			{Flack}},\ }\href {\doibase 10.1038/ncomms14301} {\bibfield  
			{journal}
		{\bibinfo  {journal} {Nature Communications}\ }\textbf {\bibinfo 
		{volume}
			{8}},\ \bibinfo {pages} {1} (\bibinfo {year} {2017})}\BibitemShut 
			{NoStop}%
	\bibitem [{\citenamefont {Koorehdavoudi}\ and\ \citenamefont
		{Bogdan}(2016)}]{koorehdavoudi2016statistical}%
	\BibitemOpen
	\bibfield  {author} {\bibinfo {author} {\bibfnamefont {H.}~\bibnamefont
			{Koorehdavoudi}}\ and\ \bibinfo {author} {\bibfnamefont 
			{P.}~\bibnamefont
			{Bogdan}},\ }\href@noop {} {\bibfield  {journal} {\bibinfo  
			{journal}
			{Scientific reports}\ }\textbf {\bibinfo {volume} {6}},\ \bibinfo 
			{pages} {1}
		(\bibinfo {year} {2016})}\BibitemShut {NoStop}%
	\bibitem [{\citenamefont {Friston}(2010)}]{friston2010free}%
	\BibitemOpen
	\bibfield  {author} {\bibinfo {author} {\bibfnamefont {K.}~\bibnamefont
			{Friston}},\ }\href@noop {} {\bibfield  {journal} {\bibinfo  
			{journal}
			{Nature reviews neuroscience}\ }\textbf {\bibinfo {volume} {11}},\ 
			\bibinfo
		{pages} {127} (\bibinfo {year} {2010})}\BibitemShut {NoStop}%
	\bibitem [{\citenamefont {Horii}\ and\ \citenamefont
		{Nemat-Nasser}(1986)}]{horii1986brittle}%
	\BibitemOpen
	\bibfield  {author} {\bibinfo {author} {\bibfnamefont {H.}~\bibnamefont
			{Horii}}\ and\ \bibinfo {author} {\bibfnamefont {S.}~\bibnamefont
			{Nemat-Nasser}},\ }\href {\doibase 10.1098/rsta.1986.0101} 
			{\bibfield
		{journal} {\bibinfo  {journal} {Philosophical Transactions of the Royal
				Society of London A: Mathematical, Physical and Engineering 
				Sciences}\
		}\textbf {\bibinfo {volume} {319}},\ \bibinfo {pages} {337} (\bibinfo 
		{year}
		{1986})}\BibitemShut {NoStop}%
	\bibitem [{\citenamefont {Byerlee}(1968)}]{byerlee1968brittle}%
	\BibitemOpen
	\bibfield  {author} {\bibinfo {author} {\bibfnamefont {J.~D.}\ \bibnamefont
			{Byerlee}},\ }\href {\doibase 10.1029/JB073i014p04741} {\bibfield  
			{journal}
		{\bibinfo  {journal} {Journal of Geophysical Research}\ }\textbf 
		{\bibinfo
			{volume} {73}},\ \bibinfo {pages} {4741} (\bibinfo {year}
		{1968})}\BibitemShut {NoStop}%
	\bibitem [{\citenamefont {Jaynes}(1957)}]{jaynes1}%
	\BibitemOpen
	\bibfield  {author} {\bibinfo {author} {\bibfnamefont {E.~T.}\ \bibnamefont
			{Jaynes}},\ }\href {\doibase 10.1103/PhysRev.106.620} {\bibfield  
			{journal}
		{\bibinfo  {journal} {Phys. Rev.}\ }\textbf {\bibinfo {volume} {106}},\
		\bibinfo {pages} {620} (\bibinfo {year} {1957})}\BibitemShut {NoStop}%
	\bibitem [{\citenamefont {Goldenfeld}(1992)}]{goldenfeld}%
	\BibitemOpen
	\bibfield  {author} {\bibinfo {author} {\bibfnamefont {N.}~\bibnamefont
			{Goldenfeld}},\ }\href@noop {} {\emph {\bibinfo {title} {Lectures 
			on phase
				transitions and the renormalization group}}}\ (\bibinfo  
				{publisher}
	{Addison-Wesley},\ \bibinfo {address} {Reading MA},\ \bibinfo {year}
	{1992})\BibitemShut {NoStop}%
	\bibitem [{\citenamefont {Withers}(2007)}]{withers2007residual}%
	\BibitemOpen
	\bibfield  {author} {\bibinfo {author} {\bibfnamefont {P.}~\bibnamefont
			{Withers}},\ }\href {\doibase 10.1088/0034-4885/70/12/R04} 
			{\bibfield
		{journal} {\bibinfo  {journal} {Reports on progress in physics}\ 
		}\textbf
		{\bibinfo {volume} {70}},\ \bibinfo {pages} {2211} (\bibinfo {year}
		{2007})}\BibitemShut {NoStop}%
	\bibitem [{\citenamefont {Li}\ \emph {et~al.}(1991)\citenamefont {Li},
		\citenamefont {Wang},\ and\ \citenamefont 
		{Backer}}]{li1991micromechanical}%
	\BibitemOpen
	\bibfield  {author} {\bibinfo {author} {\bibfnamefont {V.~C.}\ \bibnamefont
			{Li}}, \bibinfo {author} {\bibfnamefont {Y.}~\bibnamefont {Wang}}, 
			\ and\
		\bibinfo {author} {\bibfnamefont {S.}~\bibnamefont {Backer}},\ }\href
	{\doibase 10.1016/0022-5096(91)90043-N} {\bibfield  {journal} {\bibinfo
			{journal} {Journal of the Mechanics and Physics of Solids}\ }\textbf
		{\bibinfo {volume} {39}},\ \bibinfo {pages} {607} (\bibinfo {year}
		{1991})}\BibitemShut {NoStop}%
	\bibitem [{\citenamefont {Karihaloo}(1995)}]{karihaloo1995fracture}%
	\BibitemOpen
	\bibfield  {author} {\bibinfo {author} {\bibfnamefont {B.}~\bibnamefont
			{Karihaloo}},\ }\href 
			{https://books.google.com/books?id=XX1tQgAACAAJ} {\emph
		{\bibinfo {title} {Fracture Mechanics and Structural Concrete}}},\ 
		Concrete
	design and construction series\ (\bibinfo  {publisher} {Longman Scientific 
	\&
		Technical},\ \bibinfo {year} {1995})\BibitemShut {NoStop}%
\end{thebibliography}
\end{document}